\documentclass[review]{elsarticle}

\usepackage{lineno,hyperref}
\modulolinenumbers[5]
\usepackage[utf8]{inputenc}
\usepackage{bm}
\usepackage{amssymb}

\newcommand{\text}[1]{{\rm #1}}
\newcommand{\smatrix}[2]{\left#2\begin{array}{#1}}
\newcommand{\ematrix}[1]{\end{array}\right#1}

\def\schrek{Schr\"odinger}
\def\rb{\bm{r}}
\def\nb{\bm{n}}
\def\Pb{\bm{P}}
\def\Bb{\bm{B}}
\def\Rb{\bm{R}}

\def\Cb{\bm{C}}
\def\Fb{\bm{F}}
\def\taub{\bm{\tau}}

\def\difd#1{\ {\rm d}#1}
\def\der#1#2{\frac{{\rm d}#1}{{\rm d}#2}}

\def\figdir{figures}

\journal{Mathematics and Computers in Simulation}

\begin{document}

\begin{frontmatter}

\title{Isogeometric analysis in electronic structure calculations}

\author[NTC]{Robert Cimrman\corref{rc}}
\cortext[rc]{Corresponding author}
\ead{cimrman3@ntc.zcu.cz}
\author[IP]{Matyáš Novák}
\author[IT]{Radek Kolman}
\author[ICS]{Miroslav Tůma}
\author[IP]{Jiří Vackář}

\address[NTC]{New Technologies Research Centre, University of West Bohemia,
  Univerzitní 8, 306 14 Plzeň, Czech Republic}

\address[IP]{Institute of Physics, Academy of Sciences of the Czech Republic,
  Na Slovance 1999/2, Prague, Czech Republic}

\address[IT]{Institute of Thermomechanics, Academy of Sciences of the Czech
  Republic, Dolejškova 5, 182 00 Prague, Czech Republic}

\address[ICS]{Institute of Computer Science, Academy of Sciences of the Czech
  Republic, Pod Vodárenskou věží 2, 182 07, Prague, Czech Republic}

\begin{abstract}
  In electronic structure calculations, various material properties can be
  obtained by means of computing the total energy of a system as well as
  derivatives of the total energy w.r.t. atomic positions. The derivatives,
  also known as Hellman-Feynman forces, require, because of practical
  computational reasons, the discretized charge density and wave functions
  having continuous second derivatives in the whole solution domain. We
  describe an application of isogeometric analysis (IGA), a spline modification
  of finite element method (FEM), to achieve the required continuity. The
  novelty of our approach is in employing the technique of Bézier extraction to
  add the IGA capabilities to our FEM based code for ab-initio calculations of
  electronic states of non-periodic systems within the density-functional
  framework, built upon the open source finite element package SfePy. We
  compare FEM and IGA in benchmark problems and several numerical results are
  presented.
\end{abstract}


\end{frontmatter}

\linenumbers

\section{Introduction}

The electronic structure calculations allow to predict and understand material
properties such as stable atomic arrangements by minimizing the total internal
energy of a system of atoms, as well as to determine derived properties such as
elasticity, hardness, electric and magnetic properties, etc.

We are developing a real space code \cite{ptcp} for electronic structure
calculations based on
\begin{itemize}
\item the density functional theory (DFT), \cite{DFT-1, DFT-2, DFT-3, DFT-4};
\item the environment-reflecting pseudopotentials \cite{vackar};
\item a weak solution of the Kohn-Sham equations \cite{Kohn-Sham}.
\end{itemize}
The code is based on the open source finite element package SfePy (Simple
Finite Elements in Python, \url{http://sfepy.org}) \cite{SfePy-1}, which is a
general package for solving (systems of) partial differential equations (PDEs)
by the finite element method (FEM), cf.~\cite{FEM-1, FEM-2}.

The key required ability for practical computations is the calculation of the
\emph{Hellman-Feynman forces} (HFF), which correspond to the derivatives of the
total energy w.r.t. atomic positions. The HFF can efficiently provide gradients
in a gradient-based optimizer searching the total energy minimum. A major
hurdle to overcome in computing the HFF is the requirement that the discretized
charge density and wave functions should have continuous second derivatives in
the whole solution domain --- implementing a globally $C^2$ continuous basis in
FEM is not easy. Therefore we decided to enhance the SfePy package with another
PDE discretization scheme, the \emph{Isogeometric analysis} (IGA), see
\cite{IGA-1, IGA-2}.

IGA is a modification of FEM which employs shape functions of different spline
types such as B-splines, NURBS (Non-uniform rational B-spline), T-splines
\cite{IGA-6}, etc. It was successfully employed for numerical solutions of
various physical and mathematical problems, such as fluid dynamics, diffusion
and other problems of continuum mechanics~\cite{IGA-1}.

IGA has been reported to have excellent convergence properties when solving
eigenvalue problems connected to free vibrations in elasticity \cite{IGA-8},
where the errors in frequencies decrease in the whole frequency band with
increasing the approximation order, so even for high frequencies the accuracy
is very good. It should be noted that dispersion and frequency errors are
reported to decrease with increasing spline order
\cite{IGA-Waves-Hughes}. Moreover, IGA solution excellently approximates not
only eigenvalues in the full frequency spectrum but also accurately
approximates eigen-modes. There are no optical modes as in higher-order FEM,
where the errors in higher frequencies grow rapidly with the approximation
order and band gaps in the frequency band exist, see \cite{IGA-3, IGA-4,
  IGA-5}. The Kohn-Sham equations are a highly non-linear eigenvalue problem so
the above properties of IGA seem to further support our choice.

The drawbacks of using IGA, as reported also in \cite{IGA-8}, concern mainly
the increased computational cost of the numerical integration and
assembling. Also, because of the higher global continuity, the assembled
matrices have more nonzero entries than the matrices corresponding to the $C^0$
FEM basis. A comparison study of IGA and FEM matrix structures, the cost of
their evaluation, and mainly the cost of direct and iterative solvers in IGA
has been presented by \cite{IGA-Collier} and \cite{IGA-Schillinger-1}.

Recently, using FEM and its variants in electronic structure calculation
context is pursued by a growing number of groups, cf.~\cite{DFT-FEM-Davydov},
where the $hp$-adaptivity is discussed, \cite{DFT-FEM-Motamarri-1,
  DFT-FEM-Motamarri-2} where spectral finite elements as well as the
$hp$-adaptivity are considered, or \cite{DFT-IGA-Masud}, where NURBS-based FEM
is applied.

In the paper we first outline the physical problem of electronic states
calculations in Section~\ref{sec:ces}, then introduce the computational methods
and their implementation in terms of both FEM and IGA in
Section~\ref{sec:implementation}. Finally, we present a comparison of FEM and
IGA using a benchmark problem and show some numerical results in
Section~\ref{sec:examples}.

\section{Calculation of electronic states}
\label{sec:ces}

The DFT allows decomposing the many-particle \schrek{} equation into the
one-electron Kohn-Sham equations. Using atomic units they can be written in the
common form
\begin{equation}
  \label{eq:kohn-sham}
  \left(-\frac{1}{2} \nabla^2 + V_{\text{H}}(\rb) + V_{\text{xc}}(\rb)
    + \hat{V}(\rb) \right) \psi_i = \varepsilon_i \psi_i \;,
\end{equation}
which provide the orbitals $\psi_i$ that reproduce, with the weights of
occupations $n_i$, the charge density $\rho$ of the original interacting
system, as
\begin{equation}
  \label{eq:density}
  \rho(\rb) = \sum_i^N n_i |\psi_i(\rb)|^2 \;.
\end{equation}
$\hat{V}$ is a (generally) non-local Hermitian operator representing the
effective ionic potential for electrons. In the present case, within
pseudopotential approach, $\hat{V}$ represents core electrons, separated from
valence electrons, together with the nuclear charge. $V_{\text{xc}}$ is the
exchange-correlation potential describing the non-coulomb electron-electron
interactions. We use local-density
aproximation (LDA) of this potential \cite{DFT-3}. $V_{\text{H}}$ is the
electrostatic potential obtained as a solution to the Poisson equation. The
Poisson equation for $V_{\text{H}}$ has the charge density $\rho$ at its
right-hand side and is as follows:
\begin{equation}
  \label{eq:poisson}
  \Delta V_{\text{H}} = 4 \pi \rho \;.
\end{equation}
Denoting the total potential $V := V_{\text{H}} + V_{\text{xc}} + \hat{V}$, we
can write, using Hartree atomic units,
\begin{equation}
  \label{eq:kohn-sham2}
  \left(-\frac{1}{2} \nabla^2 + V(\rb) \right) \psi_i = \varepsilon_i \psi_i \;.
\end{equation}
Note that the above mentioned eigenvalue problem is highly non-linear, as the
potential $V$ depends on the orbitals $\psi_i$. Therefore an iterative scheme
is needed, defining the DFT loop for attaining a self-consistent solution.

\subsection{DFT loop}

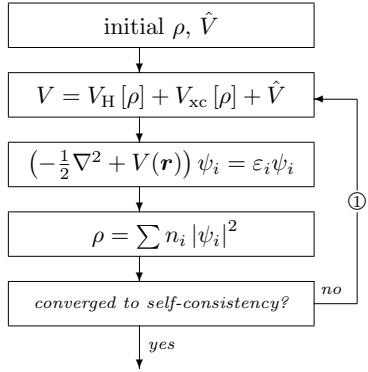
\begin{figure}[h]
  \vspace*{20pt}
  \begin{center}
    \setlength{\unitlength}{0.85ex}
    \scalebox{0.9}{
      \begin{picture}(60,50)
        \put(10,50){\framebox(35,5){initial $\rho$, $\hat{V}$}}
        \put(25,50){\vector(0,-1){3}}
        \put(10,42){\framebox(35,5){$V = V_{\text{H}} \left [ \rho \right ]
            + V_{\text{xc}} \left [ \rho \right ] + \hat{V}$}}
        \put(25,42){\vector(0,-1){3}}
        \put(10,34){\framebox(35,5){$\left(-\frac{1}{2} \nabla^2
              + V(\rb) \right) \psi_i = \varepsilon_i \psi_i$}}
        \put(25,34){\vector(0,-1){3}}
        \put(10,26){\framebox(35,5){
            $\rho = \sum n_i \left | \psi_i \right |^2$}}
        \put(25,26){\vector(0,-1){3}}
        \put(10,18){\framebox(35,5){\scriptsize \it converged to
            self-consistency?}}

        \put(45,20.5){\line(1,0){5}}
        \put(50,20.5){\line(0,1){10.5}}
        \put(50,32.25){\circle{2}}
        \put(50,32.25){\makebox(0,0){\scriptsize 1}}
        \put(50,33.5){\line(0,1){10.5}}
        \put(50,44.0){\vector(-1,0){5}}

        \put(45,20.5){\makebox(4,3){\scriptsize \it no}}
        \put(25,18){\vector(0,-1){5}}
        \put(25,14.0){\makebox(5,3){\scriptsize \it yes}}
      \end{picture}
    }
  \end{center}
  \vspace*{-20mm}
  \caption{DFT, iterative self-consistent scheme.} \label{fig:scheme}
\end{figure}

For the global convergence of the DFT iteration we use the standard algorithm
outlined in Fig.~\ref{fig:scheme}. The purpose of the DFT loop is to find a
self-consistent solution. Essentially, we are seeking a fixed point of a
function of $V_{\text{hxc}} := V_{\text{H}} \left [ \rho \right ] +
V_{\text{xc}} \left [ \rho \right ]$. For this task, a variety of nonlinear
solvers can be used. We use Broyden-type quasi-Newton solvers applied to
\begin{equation}
  \label{eq:dft}
  DFT(V_{\text{hxc}}^{i}) - V_{\text{hxc}}^{i} = V_{\text{hxc}}^{i+1} -
  V_{\text{hxc}}^{i} = 0 \;,
\end{equation}
where $DFT$ denotes a single iteration of the DFT loop.

After the DFT loop convergence is achieved, the derived quantities,
particularly the total energy, are computed. By minimizing the total energy as
a function of atomic positions, the equilibrium atomic positions can be found.
Therefore the DFT loop itself can be embedded into an outer optimization loop,
where the objective function gradients are the HFF.

\subsection{Forces acting on atoms}

The force acting on atom $\alpha$ at position $\taub_\alpha$ is equal to the
derivative of the total energy functional with respect to an infinitesimal
displacement of this atom $\delta\taub_\alpha$:
\begin{equation}
  \label{eq:force1}
  \Fb^\alpha = - \frac{\delta E}{\delta\taub_\alpha} \;.
\end{equation}

Making use of the Hellmann-Feynman theorem that relates the derivative of the
total energy, with respect to a parameter $\lambda$, to the expectation value of
the derivative of the Hamiltonian operator, w.r.t. the same parameter,
\begin{equation}
  \label{eq:hft}
  \der{E}{\lambda}
  = \left\langle \Psi^{*}_{\lambda}
  \left| \der{\hat{H}_\lambda}{\lambda} \right|
  \Psi_\lambda \right\rangle \;,
\end{equation}
within the density functional theory we can write
\begin{equation}
  \label{eq:force2}
  \Fb^\alpha = \Fb^\alpha_{\rm HF,es} -
  \frac{\left( \sum_i n_i \delta\varepsilon_i -
      \int \rho(\rb) \delta V_{\rm eff} (\rb) d^3 \rb
    \right)}{\delta \taub_\alpha} \;,
\end{equation}
where the first term is the electrostatic Hellmann-Feynman force and the second
term is the ``Pulay'' force, also known as ``incomplete basis set'' force,
that contains the corrections that depend on technical details of the
calculation. The electrostatic Hellmann-Feynman force is given by the sum over
all the atoms $\beta\neq\alpha$ of electrostatic forces between the charges of
atomic nuclei $Z_\alpha$ and $Z_\beta$ and by the force acting on the charge
$Z_\alpha$ in the electric field of the charge density $\rho$:
\begin{equation}
  \label{eq:hff-es}
  \Fb^\alpha_{\rm HF,es} = Z_\alpha \der{}{\taub_\alpha}
  \left(
    - \sum_{\beta\neq\alpha} \frac{Z_\beta}{|\taub_\alpha - \taub_\beta|}
    + \int \frac{\rho(\rb)}{|\taub_\alpha-\rb|} d^3 \rb
  \right) \;.
\end{equation}
The effective potential
\begin{equation}
  \label{eq:hff-veff}
  V_{\rm eff}=-\sum_\alpha \frac{Z_\alpha}{|\rb - \taub_\alpha|} +
  V_{\rm hxc}(\rb)
\end{equation}
used in the Pulay force term, together with the kinetic energy operator, forms
the total energy (for more details see e.g. \cite{Krakauer},
\cite{Weinert}, \cite{Alessandra}).

Within the pseudopotential formalism, as it was shown by Ihm, Zunger and Cohen
\cite{IhmZungerCohen}, the electrostatic HF force (\ref{eq:hff-es}) transforms
into
\begin{equation}
  \label{eq:es-izc}
  - \sum_{\beta\neq\alpha}
  \der{
  }{\taub_\alpha}
    \left(
      \frac{Z_\alpha Z_\beta}{|\taub_\alpha - \taub_\beta|}
    \right)
  +
  \sum_l \int \rho_{{\rm ion},l}(|\rb
  - \taub_\alpha|)E_l(\rb) d^3 \rb \;,
\end{equation}
where $\rho_{{\rm ion},l}$ is ``virtual ionic partial charge'' derived from the
$l$\/-component $U_{{\rm ps},l}$ of the semilocal form of the pseudopotential,
\begin{equation}
  \label{eq:es-izc-s1}
  \rho_{{\rm ion},l}(\rb)
  =-\frac{\pi}{4}\nabla^2 U_{{\rm ps},l}(\rb) \;,
\end{equation}
and
\begin{equation}
  \label{eq:es-izc-s2}
  E_l(\rb) = - \nabla_{\rb}
  \int \frac{\rho_{{\rm ps},l}(\rb^\prime)}{|\rb^\prime
    - \rb|}d^3 \rb^\prime \;.
\end{equation}
Here $\rho_{{\rm ps},l}$ is the $l$\/-projected (via an integration over angles
on a unit sphere) charge w.r.t. atomic center $\alpha$
\begin{equation}
  \label{eq:es-izc-s3}
  \rho_{{\rm ps},l}(\rb^\prime)=\sum_i n_i \int \psi^{*}_i
  (\rb^\prime) \hat{P}_l^\alpha \psi_i (\rb^\prime) d\theta d\phi \;,
\end{equation}
where $\hat{P}_l^\alpha$ is the Legendre-polynomial-based projector to the
$l$\/-subspace w.r.t. site $\alpha$.

The continuity of the derivatives of the wave functions $\psi_i$ up to the
second order is the necessary condition for the validity of the derivation
above; otherwise everything would become much more complicated and unsuitable
for practical use, in connection with FEM/IGA approach (for more details see
e.g. \cite{Krakauer}, \cite{Weinert}, \cite{Alessandra}).

\section{Computational methods and their implementation}
\label{sec:implementation}

\subsection{Weak formulation}
\label{sec:weak}

Let us denote $H^1(\Omega)$ the usual Sobolev space of functions with $L^2$
integrable derivatives and $H^1_0(\Omega) = \{u \in H^1(\Omega)
| u = 0 \mbox{ on } \partial \Omega\}$.

The eigenvalue problem (\ref{eq:kohn-sham2}) can be rewritten using the weak
formulation: find functions $\psi_i \in H^1(\Omega)$ such that for all $v \in
H^1_0(\Omega)$ holds
\begin{equation}
  \label{eq:kohn-sham-weak}
  \int_\Omega \frac{1}{2} \nabla \psi_i \cdot \nabla v \difd{V}
  + \int_\Omega v V \psi_i \difd{V}
  = \varepsilon_i \int_\Omega v \psi_i \difd{V}
  + \oint_{\partial \Omega} \frac{1}{2} \der{\psi_i}{\nb} \difd{S} \;.
\end{equation}
If the solution domain $\Omega$ is sufficiently large, the last term
can be neglected. The Poisson equation (\ref{eq:poisson}) has the following
weak form:
\begin{equation}
  \label{eq:poisson-weak}
  \int_{\Omega} \nabla v \cdot \nabla V_{\text{H}} = 4 \pi \int_{\Omega} \rho v \;.
\end{equation}

Equations (\ref{eq:kohn-sham-weak}), (\ref{eq:poisson-weak}) then need to be
discretized --- the continuous fields are approximated by discrete fields with
a finite set of degrees of freedom (DOFs) and a basis, typically piece-wise
polynomial:
\begin{equation}
  \label{eq:discretize}
  u(\rb) \approx u^h(\rb) = \sum_{k=1}^{N} u_k \phi_k(\rb)
  \mbox{ for } \rb \in \Omega \;,
\end{equation}
where $u$ is a continuous field ($\psi$, $v$, $V_H$ in our equations), $u_k$,
$k = 1, 2, \dots, N$ are the discrete DOFs and $\phi_k$ are the basis
functions. From the computational point of view it is desirable that the basis
functions have a small support, so that the resulting system matrix is sparse.

\subsection{Finite element method}
\label{sec:fem}

In the FEM the discretization process involves the discretization of the domain
$\Omega$ --- it is replaced by a polygonal domain $\Omega_h$ that is covered by
small non-overlapping subdomains called \emph{elements} (e.g. triangles or
quadrilaterals in 2D, tetrahedrons or hexahedrons in 3D), cf.~\cite{FEM-1,
  FEM-2}. The elements form a FE \emph{mesh}.

The basis functions are defined as piece-wise polynomials over the individual
elements, have a small support and are typically globally $C^0$ continuous. The
discretized equations are evaluated over the elements as well to obtain local
matrices or vectors that are then assembled into a global sparse system. The
evaluation usually involves a numerical integration on a reference element, and
a mapping to individual physical elements \cite{FEM-1, FEM-2}. The nodal basis
of Lagrange interpolation polynomials or the hierarchical basis of Lobatto
polynomials can be used in our code.

\subsection{Isogeometric analysis}
\label{sec:iga}

In IGA, the CAD geometrical description in terms of NURBS patches is used
directly for the approximation of the unknown fields, without the intermediate
FE mesh --- the meshing step is removed, which is one of its principal
advantages. A D-dimensional geometric domain can be defined by
\begin{equation}
  \label{eq:iga-1}
  \rb(\underline{\xi})
  = \sum_{A=1}^{n} \Pb_A R_{A,p}(\underline{\xi})
  = \Pb^T \Rb(\underline{\xi}) \;,
\end{equation}
where $\underline{\xi} = \{\xi_1, \dots, \xi_D\}$ are the parametric
coordinates, and $\Pb = \{\Pb_A\}_{A=1}^{n}$ is the set of control
points, $R_{A,p}$, $A = 1, 2, \dots n$ are the NURBS solid basis functions and
$p$ is the NURBS solid degree.

If $D > 1$, the NURBS solid can be defined as a tensor product of univariate
NURBS curves. First a mapping is defined, see \cite{IGA-2}, that maps between
the tensor product space and the global indexing of the basis functions. Let $i
= 1, 2, \dots, n$, $j = 1, 2, \dots, m$ and $k = 1, 2, \dots, l$, then
\begin{displaymath}
  \tilde{A}(i, j, k) = (l \times m)(i - 1) + l(j-1) + k \;.
\end{displaymath}
If $N_{i, p}(\xi)$, $M_{j, q}(\eta)$ and $L_{k, r}(\zeta)$ are the univariate
B-spline basis functions with degrees $p$, $q$ and $r$, respectively, then
with $A = \tilde{A}(i, j, k)$ and $\hat{A} = \tilde{A}(\hat{i}, \hat{j},
\hat{k})$
\begin{displaymath}
  R_A^{p,q,r}(\xi, \eta, \zeta) = \frac{L_{i, r}(\zeta)M_{j, q}(\eta)N_{k,
      p}(\xi) w_A}{\sum_{\hat{i}=1}^n\sum_{\hat{j}=1}^m\sum_{\hat{k}=1}^l
    L_{\hat{i}, r}(\zeta) M_{\hat{j}, q}(\eta) N_{\hat{k},p}(\xi) w_{\hat{A}}}
\end{displaymath}
are the NURBS solid basis functions (here $R_A^{p,q,r}$ corresponds to
$R_{A,p}$ in (\ref{eq:iga-1})), and $w_A$ are the weights (products of
univariate NURBS basis weights). Below we will denote $(\xi, \eta, \zeta)$ by a
vector $\underline{\xi}$. The univariate B-spline basis functions are defined
by a \emph{knot vector}, that is the vector of non-decreasing parametric
coordinates $\Xi = \{\xi_1, \xi_2, \dots, \xi_{n + p + 1}\}$, where $\xi_A \in
\mathbb{R}$ is the $A^{th}$ knot and $p$ is the polynomial degree of the
B-spline basis functions \cite{NURBS}. Then for $p = 0$
\begin{displaymath}
  \label{eq:b-spline-basis}
  N_{A,0}(\xi) =
  \smatrix{c}{\{}
  1 \mbox{ for } \xi_A \leq \xi < \xi_{A+1} \;, \\
  0 \mbox{ otherwise.}
  \ematrix{.}
\end{displaymath}
For $p > 0$ the basis functions are defined by the Cox-de Boor recursion
formula (defining $\frac{0}{0} \equiv 0$)
\begin{displaymath}
  N_{A,p}(\xi) = \frac{\xi - \xi_A}{\xi_{A+p} - \xi_A} N_{A,p-1}(\xi)
  + \frac{\xi_{A+p+1} - \xi}{\xi_{A+p+1} - \xi_{A+1}}N_{A+1,p-1}(\xi) \;.
\end{displaymath}
Note that it is possible to insert knots into a knot vector without changing
the geometric or parametric properties of the curve by computing the new set of
control points by the knot insertion algorithm, see e.g.~\cite{IGA-2}. The
continuity of approximation does not change when inserting control points. The
basic properties of the B-spline basis functions can be found in \cite{NURBS}.

In IGA, the same NURBS basis, that is used for the geometry description, is
used also for the approximation of PDE solutions. For our equation
(\ref{eq:kohn-sham-weak}) we have
\begin{equation}
  \label{eq:iga-2}
  \psi(\underline{\xi}) \approx \psi^h(\underline{\xi})
  = \sum_{A=1}^{n} \psi_A R_{A,p}(\underline{\xi})\;,
  \quad
  v(\underline{\xi}) \approx v^h(\underline{\xi})
  = \sum_{A=1}^{n} v_A R_{A,p}(\underline{\xi})\;,
\end{equation}
where $\psi_A$ are the unknown DOFs - coefficients of the basis in the linear
combination, and $v_A$ are the test function DOFs.

Complex geometries cannot be described by a single NURBS outlined above, often
called \emph{NURBS patch} --- many such patches might be needed, and special
care must be taken to ensure required continuity along patch boundaries and to
avoid holes. Usually, the patches are connected using $C^0$ continuity only, as
the individual patches have the open knot vectors \cite{NURBS}.

However, on a single patch, the NURBS basis can be smooth as needed for the HFF
calculation --- a degree $p$ curve has $p-1$ continuous derivatives, if no
internal knots are repeated, as follows from the B-spline basis properties
\cite{NURBS}. The basis functions $R_{A,p}$, $A = 1, \dots, n$ on the patch are
uniquely determined by the knot vector for each axis, and cover the whole
patch. Due to our continuity requirements, only single-patch domains are
considered in this paper. Also, we set $w_A = 1$, $A = 1, \dots, n$, thus using
a B-spline basis instead of full NURBS basis, because our domain is simply a
cube, see Section~\ref{sec:examples}.

\subsection{IGA implementation in SfePy}
\label{sec:iga-sfepy}

\begin{figure}[htp!]
  \centering
  \includegraphics[width=\linewidth]{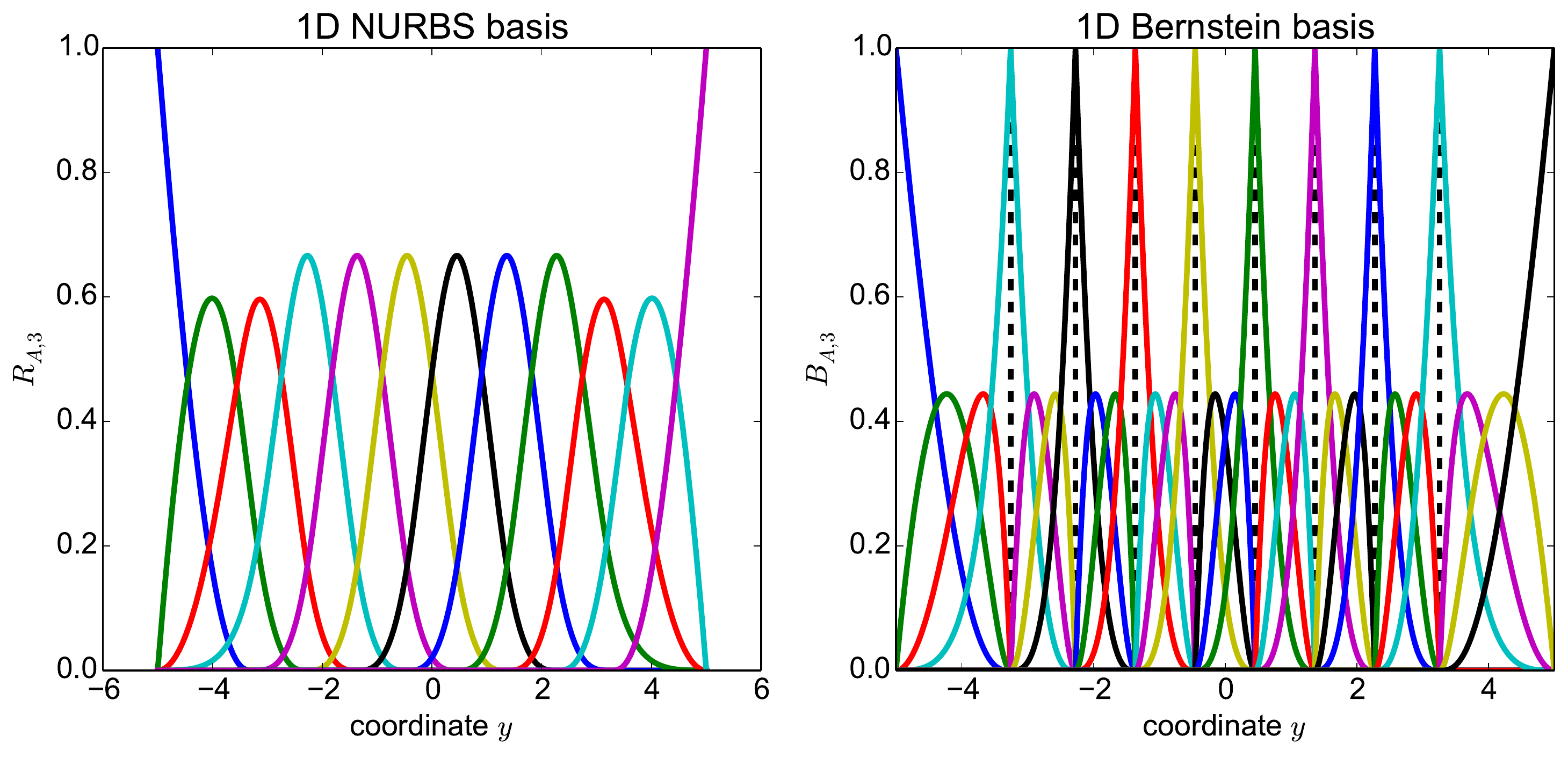}

  \caption{Left: NURBS basis of degree 3 that describes the second axis of the
    parametric mesh in Fig.~\ref{fig:ig-domain-grids}. Right: the corresponding
    Bernstein basis with Bézier elements delineated by vertical lines.}
 \label{fig:bezier-extraction}
\end{figure}

Our implementation \cite{SfePy-2} uses a variant of IGA based on \emph{Bézier
  extraction operators} \cite{IGA-2} that is suitable for inclusion into
existing FE codes. The code itself does not see the NURBS description at
all. It is based on the observation that repeating a knot in the knot vector
decreases continuity of the basis in that knot by one. This can be done in such
a way that the overall shape remains the same, but the "elements" appear
naturally as given by non-zero knot spans. The final basis restricted to each
of the elements is formed by the Bernstein polynomials $\Bb$,
cf.~\cite{IGA-2}. The Bézier extraction process is illustrated in
Fig.~\ref{fig:bezier-extraction}. The depicted basis corresponds to the second
parametric axis of the domain shown in Fig.~\ref{fig:ig-domain-grids}, see
below.

In \cite{IGA-2} algorithms are developed that allow computing the Bézier
extraction operator $\Cb$ for each element such that the original
(smooth) NURBS basis function $\Rb$ can be recovered from the local
Bernstein basis $\Bb$ using $\Rb = \Cb\Bb$. The Bézier extraction
also allows construction of the Bézier mesh, see
Fig.~\ref{fig:bezier-extraction}, right and Fig.~\ref{fig:ig-domain-grids},
right. The code then loops over the Bézier elements and assembles local
contributions in the usual FE sense. The operator $\Cb$ is a function of the
knot vectors only --- it does not depend on the positions of control points.

Several kinds of grids (or ``meshes'') can be constructed for a NURBS patch, as
depicted in Fig.~\ref{fig:ig-domain-grids}.  The parametric mesh is simply the
tensor product of the knot vectors defining the parametrization --- the lines
correspond to the knot vector values. The control mesh has vertices given by
the NURBS patch control points and connectivity corresponding to the tensor
product nature of the patch. The Bézier mesh has been introduced above and its
vertices are the control points of the individual Bézier elements. In our code
we use the corner vertices of the Bézier elements to construct a
\emph{topological Bézier mesh}, which can be used for subdomain selection
(e.g. parts of the boundary, where boundary conditions need to be applied),
because its vertices are interpolatory, i.e., they are in the NURBS domain or
on its boundary.

In our implementation, full Gauss quadrature rules with the 1D quadrature order
$r = p + 1$ are used to integrate over the Bézier elements.

\begin{figure}[htp!]
  \centering
  \includegraphics[width=\linewidth]{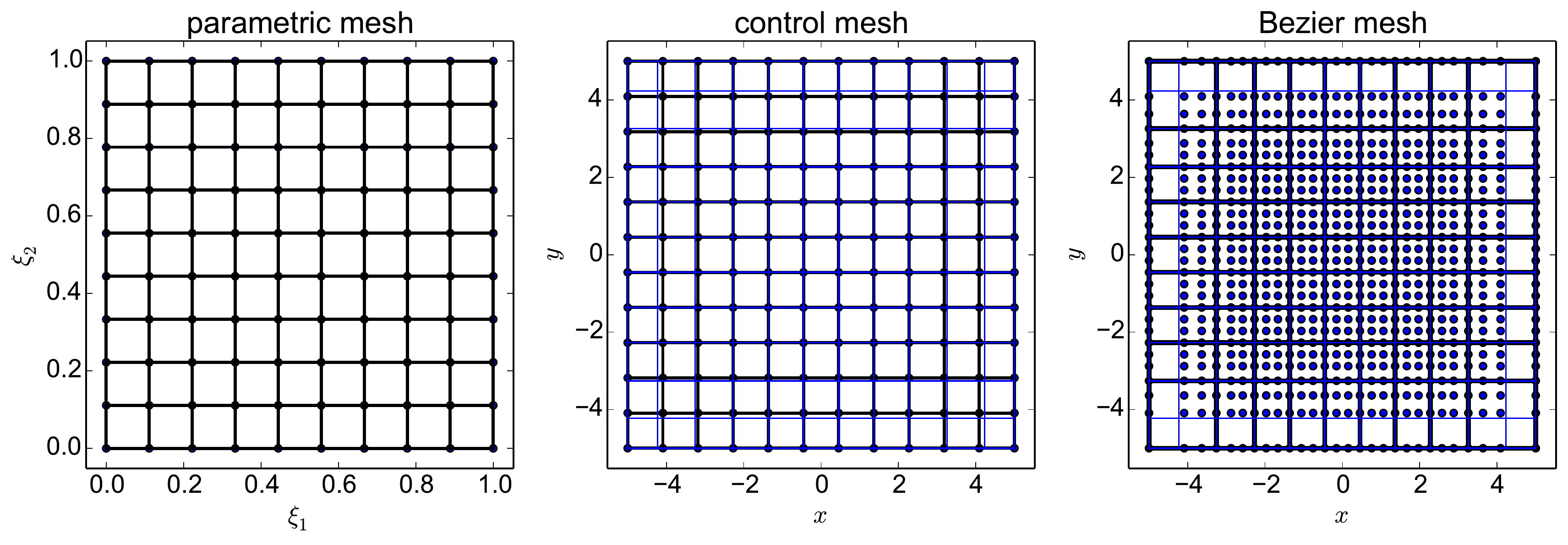}

  \caption{From left to right: parametric mesh (tensor product of knot
    vectors), control mesh, Bézier mesh. The corner vertices of Bézier mesh
    elements form the topological Bézier mesh. The thin blue lines are
    iso-lines of the NURBS parametrization.}
 \label{fig:ig-domain-grids}
\end{figure}

\section{Numerical examples}
\label{sec:examples}

In this section we show some results based on our initial tests with the IGA
implementation.

\subsection{Nitrogen atom benchmark}
\label{sec:atom}

The nitrogen atom serves us as a benchmark problem. A cube domain with the size
of $10 \times 10 \times 10$ atomic units was used for all the computations. The
FEM approximation used Lagrange polynomial basis of order three (tri-cubic
Lagrange polynomials) on a uniform hexahedral mesh. The IGA approximation used
degree 3 B-splines and a uniform knot vector in each parametric axis. The
control points were also spaced uniformly, so that the placement of the basis
in the physical space was not coarser in the middle of the cube, see
Fig.~\ref{fig:ig-domain-grids}.

\begin{figure}[ht!]
  \centering
  \includegraphics[width=0.48\linewidth]{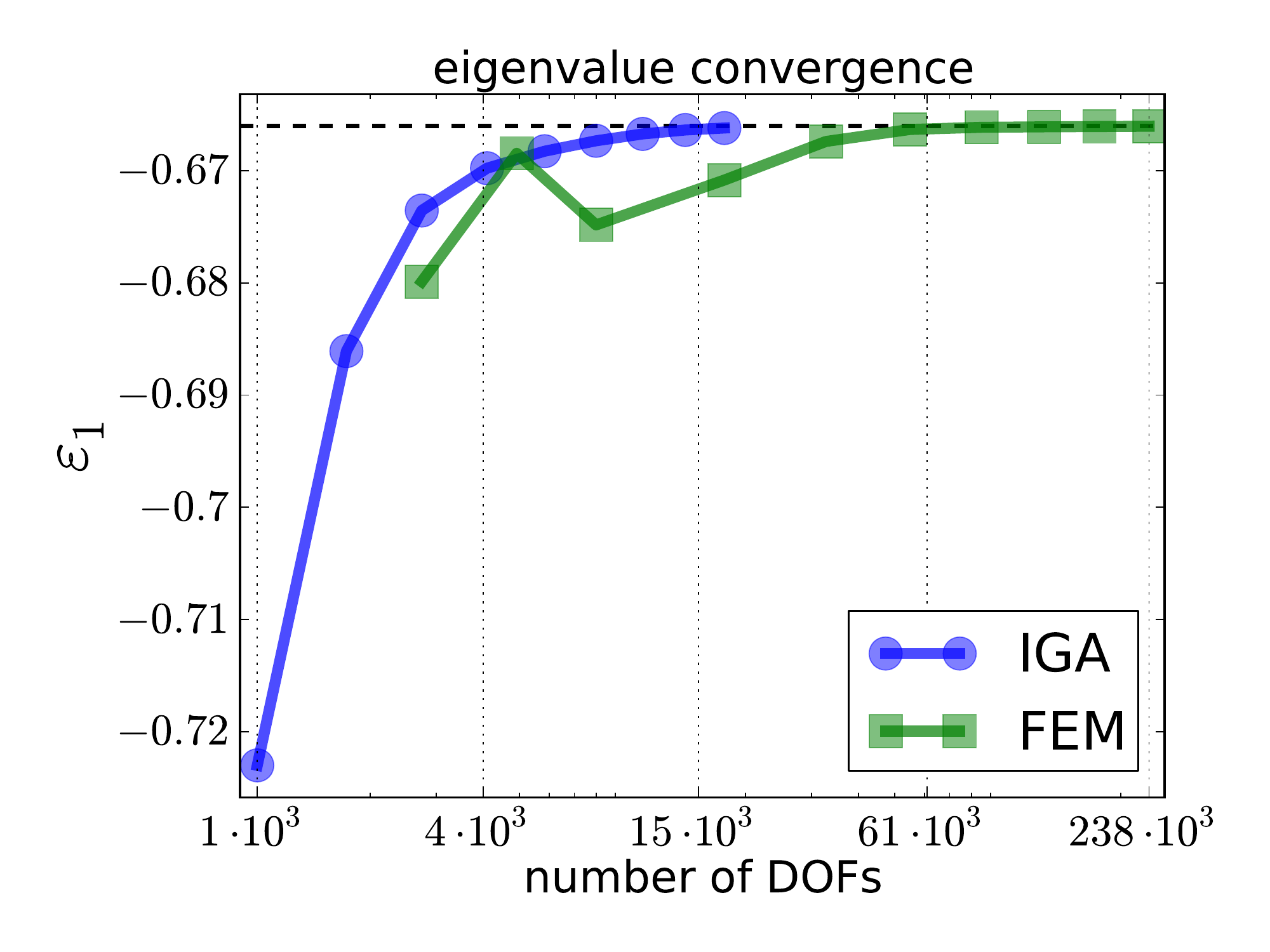}
  \hfill
  \includegraphics[width=0.48\linewidth]{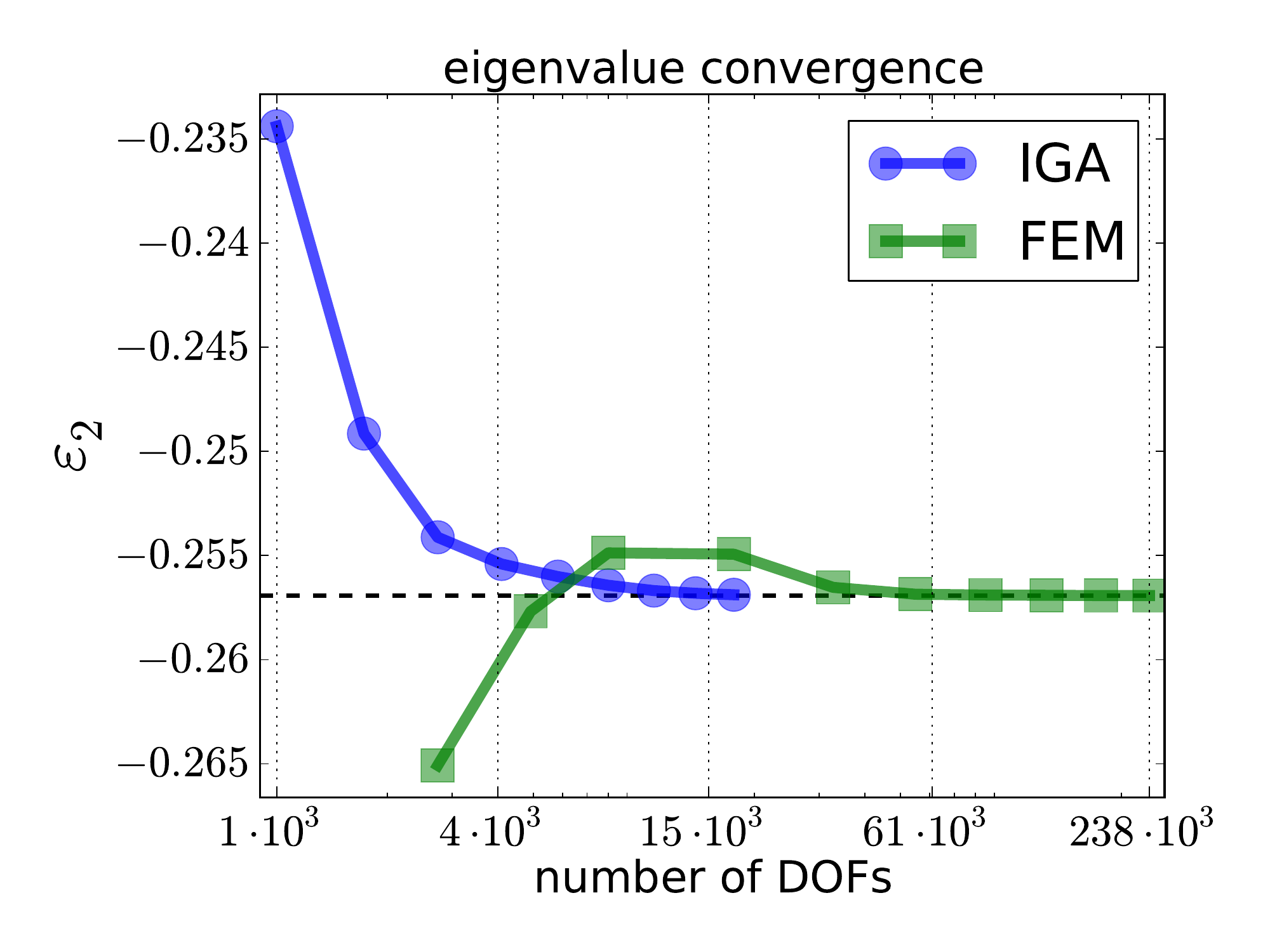}

  \caption{Convergence of eigenvalues $\varepsilon_1$ and $\varepsilon_2$
    w.r.t. the number of DOFs.}
  \label{fig:atom-conv-size}
\end{figure}

\begin{figure}[ht!]
  \centering
  \includegraphics[width=0.48\linewidth]{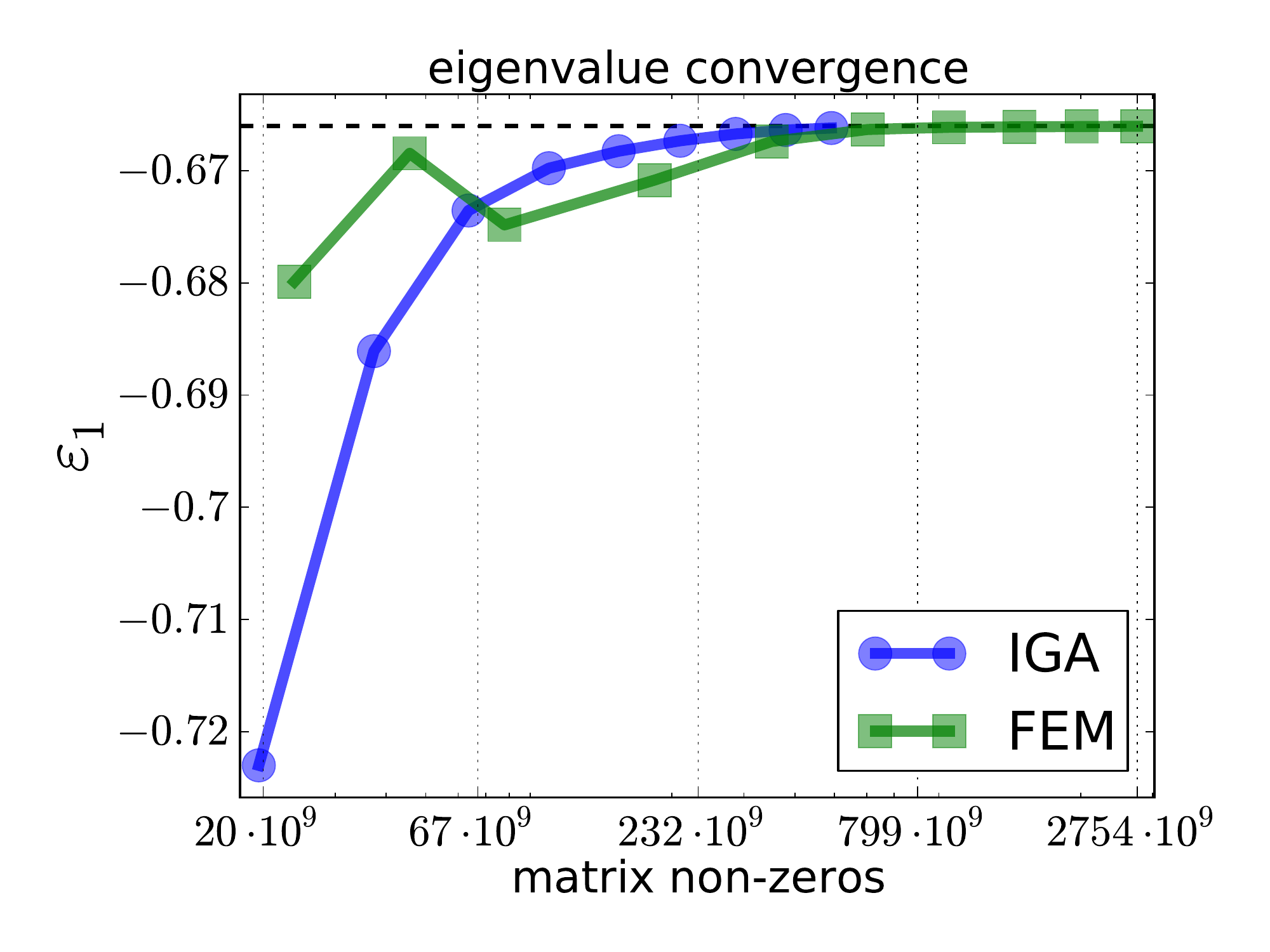}
  \hfill
  \includegraphics[width=0.48\linewidth]{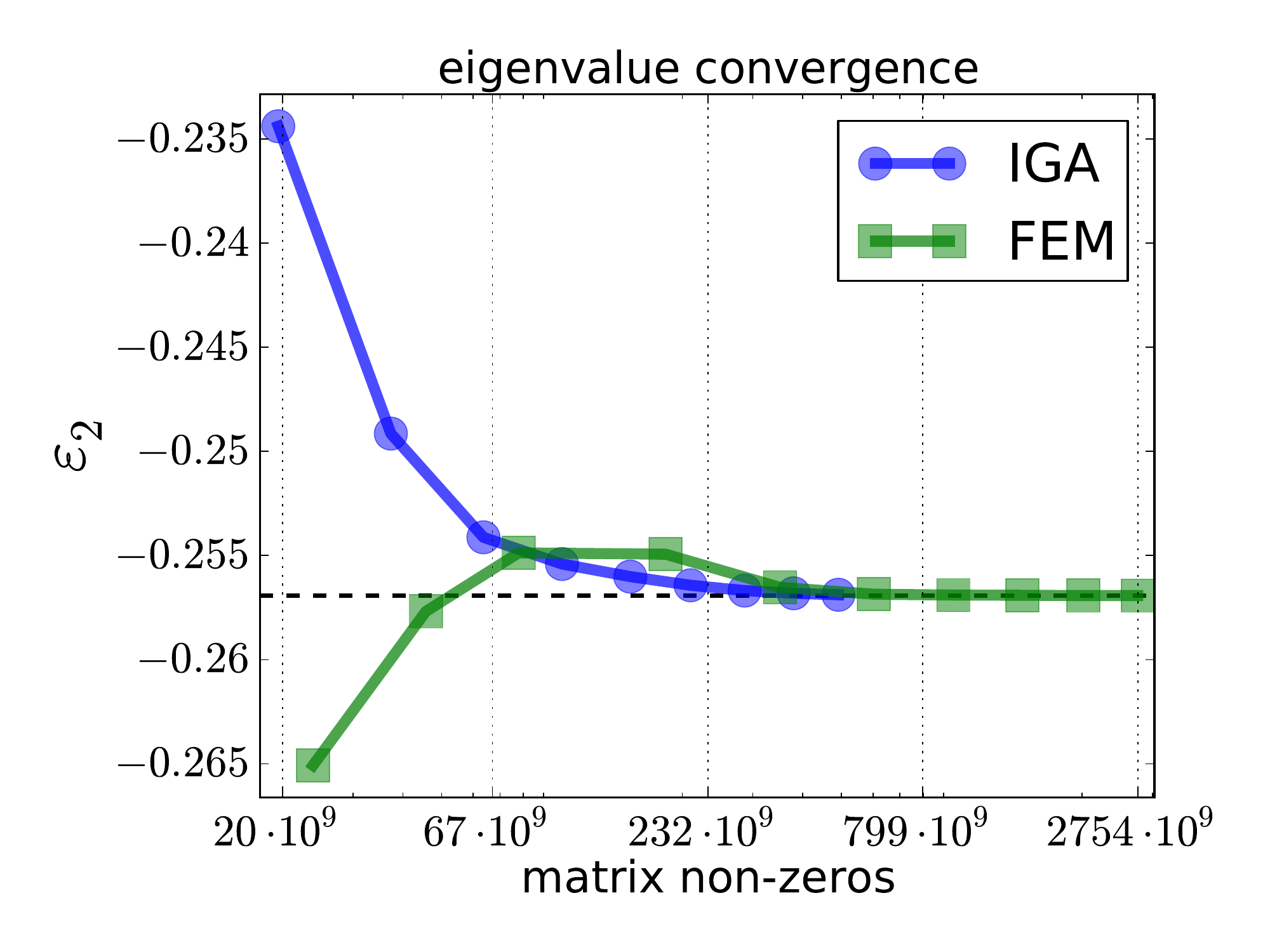}

  \caption{Convergence of eigenvalues $\varepsilon_1$ and $\varepsilon_2$
    w.r.t. the number of non-zeros of the matrices.}
  \label{fig:atom-conv-nnz}
\end{figure}

We compared IGA and FEM solution convergence given the increasing number of
DOFs (grid size). The number of DOFs corresponds to the sizes of the matrices
that come from the FEM- or IGA- discretized (\ref{eq:kohn-sham-weak}), formally
written as
\begin{equation}
  \label{eq:discrete-kohn-sham}
  (\bm{K} + \bm{V}(\bm{\psi}_i)) \bm{\psi}_i = \varepsilon_i \bm{M} \bm{\psi}_i
  \;.
\end{equation}
The following numbers of DOFs per cube side (including DOFs fixed by boundary
conditions) were used:
\begin{itemize}
\item FEM: 16, 19, 22, 28, 34, 40, 46, 52, 58, 64;
\item IGA: 12, 14, 16, 18, 20, 22, 24, 26, 28.
\end{itemize}
This corresponds to:
\begin{itemize}
\item FEM: 5,  6,  7,  9, 11, 13, 15, 17, 19, 21 cubic elements per cube side;
\item IGA: 9, 11, 13, 15, 17, 19, 21, 23, 25 Bézier elements per cube side.
\end{itemize}
The grids corresponding to the coarsest IGA approximation used are shown (2D
projection) in Fig.~\ref{fig:ig-domain-grids}. Because the difficulty of
solving (\ref{eq:discrete-kohn-sham}) depends not only on the number of DOFs,
but also on the number of non-zeros in the sparse matrices, we show the
convergence with respect to both of these parameters, see
Figs.~\ref{fig:atom-conv-size},~\ref{fig:atom-conv-nnz}. The non-zeros are
determined structurally by the compact support of each basis function and the
pattern (allocated space) is the same for both $(\bm{K} + \bm{V}(\bm{\psi}_i))$
and $\bm{M}$.

It should be noted that the converged values are neither exact physical binding
energies of electrons nor the ionization energies. They are just the Kohn-Sham
eigenvalues for a given particular problem under given conditions and
approximations, in our case with a relatively small physical domain size, since
the aim was to test quickly the numerical properties of different bases.
However, note that both methods converge to exactly the same values, which
confirms the numerical validity of the FEM/IGA calculation.

The above results suggest that IGA converges to a solution both for a much
smaller number of DOFs and for smaller number of non-zeros in the
matrices. This is in agreement with the initial prognosis that the higher-order
smooth basis functions can improve the convergence and accuracy of the finite
element electronic structure calculations. Note the non-oscillating convergence
of the IGA values in contrast to the oscillating convergence of FEM values.

\subsection{Examples of computed quantities}
\label{sec:molecule}

\begin{figure}[ht!]
  \centering

  \begin{tabular}{cccccc}
    12 &
    \includegraphics[width=0.18\linewidth]{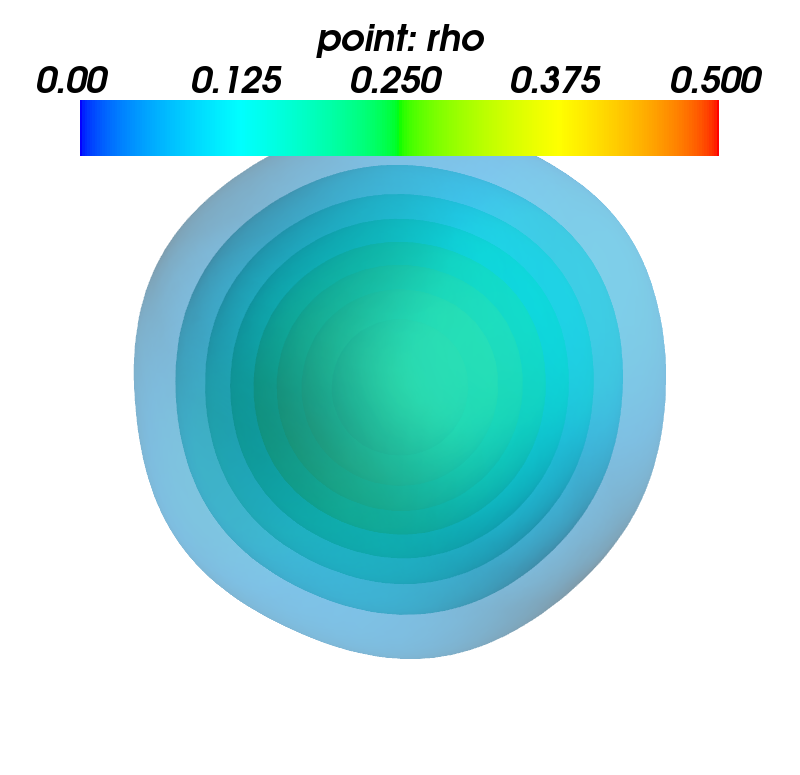}
    &
    \includegraphics[width=0.18\linewidth]{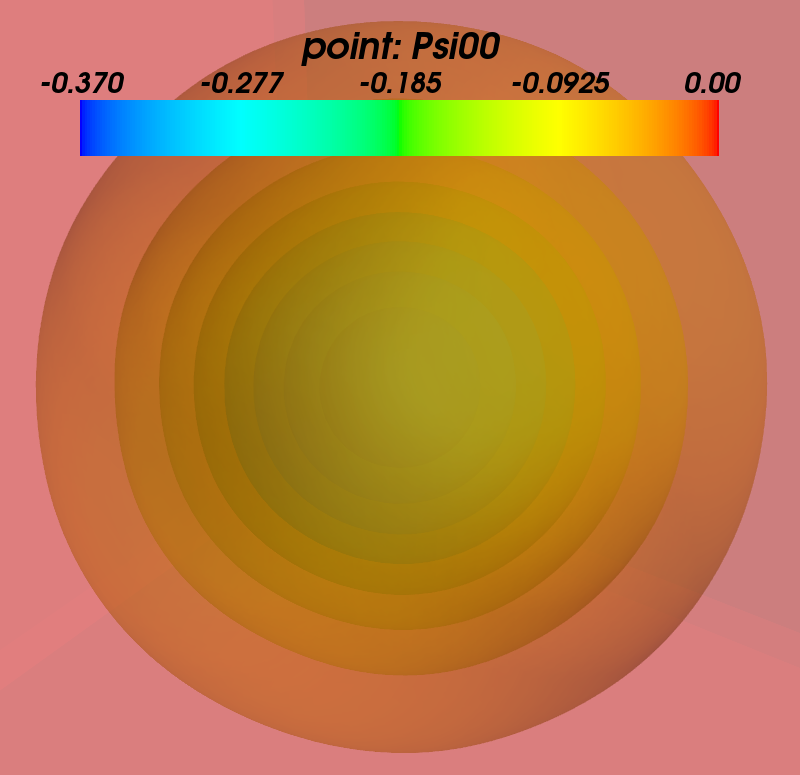}
    &
    \includegraphics[width=0.18\linewidth]{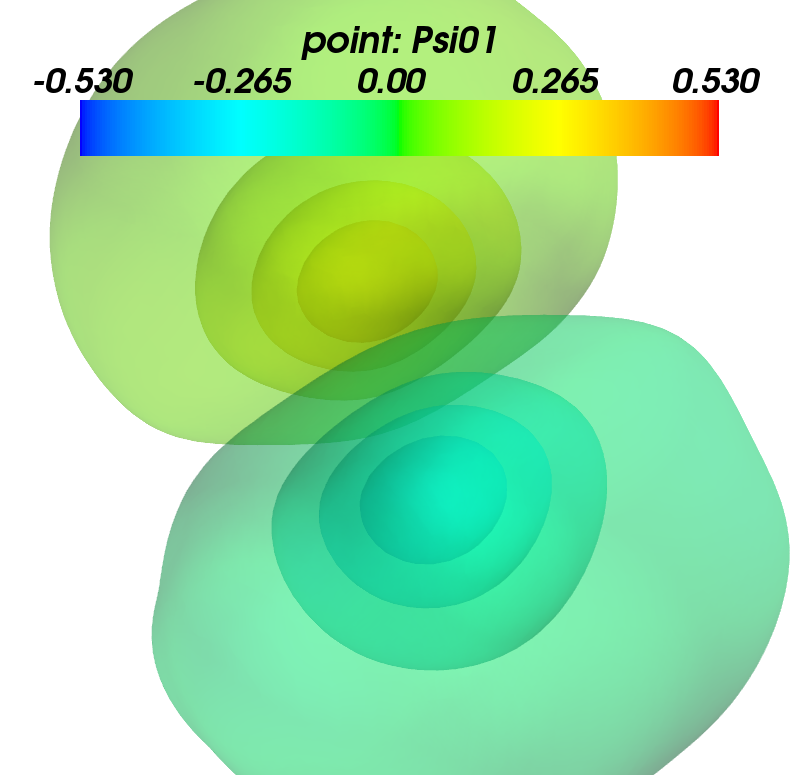}
    &
    \includegraphics[width=0.18\linewidth]{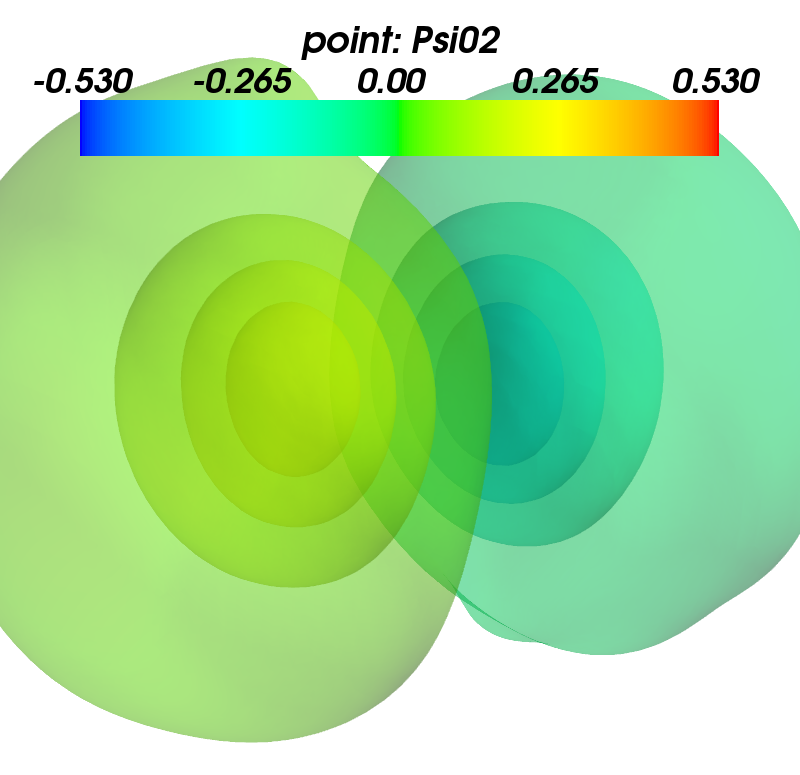}
    &
    \includegraphics[width=0.18\linewidth]{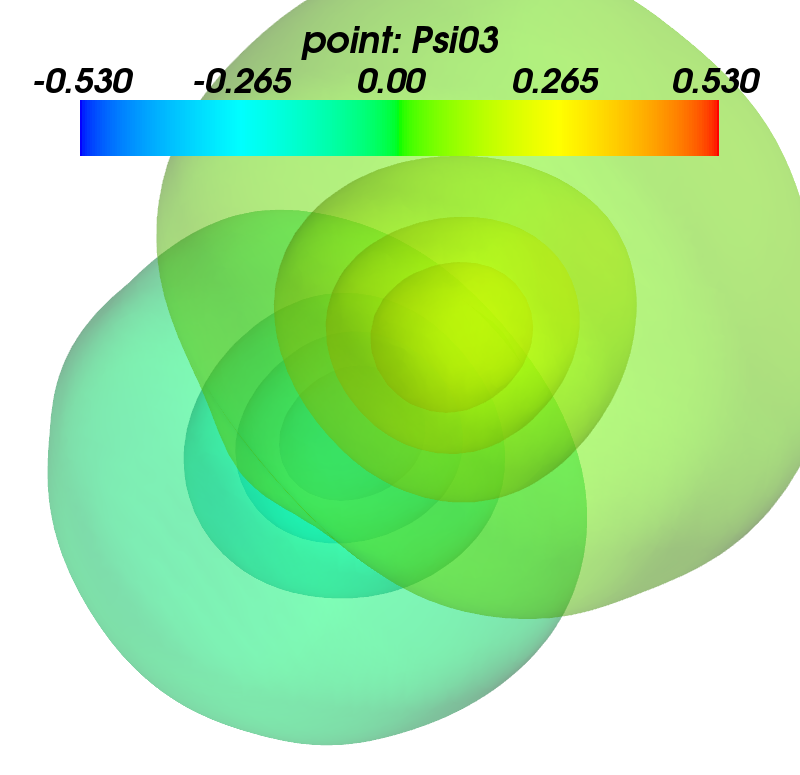}
    \\
    16 &
    \includegraphics[width=0.18\linewidth]{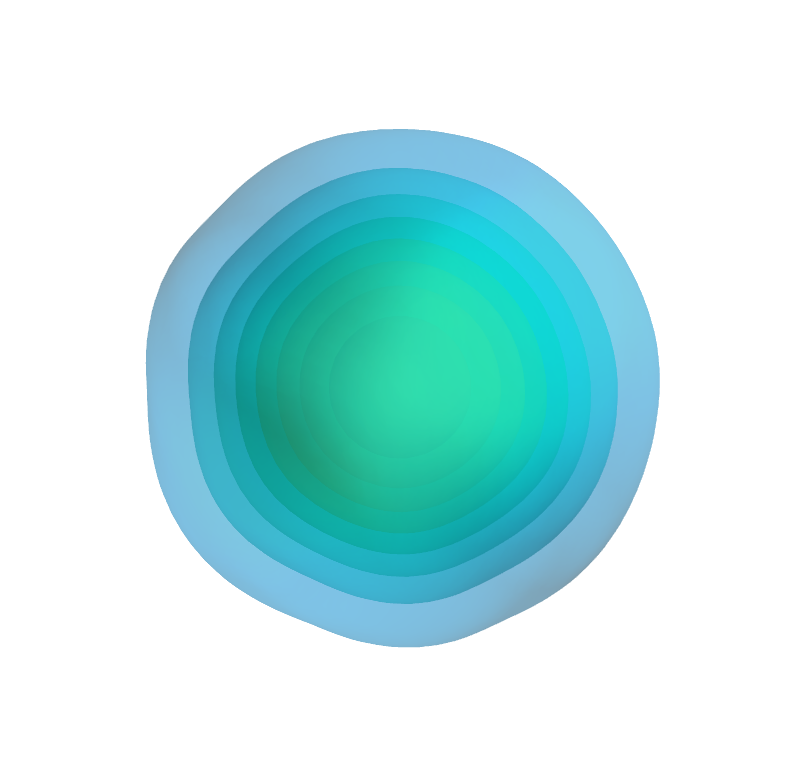}
    &
    \includegraphics[width=0.18\linewidth]{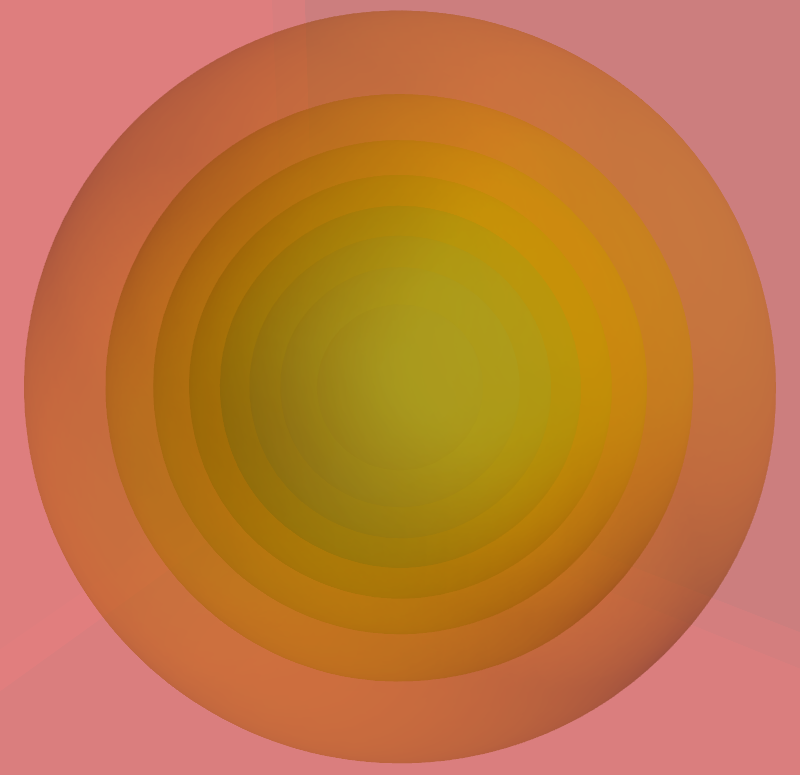}
    &
    \includegraphics[width=0.18\linewidth]{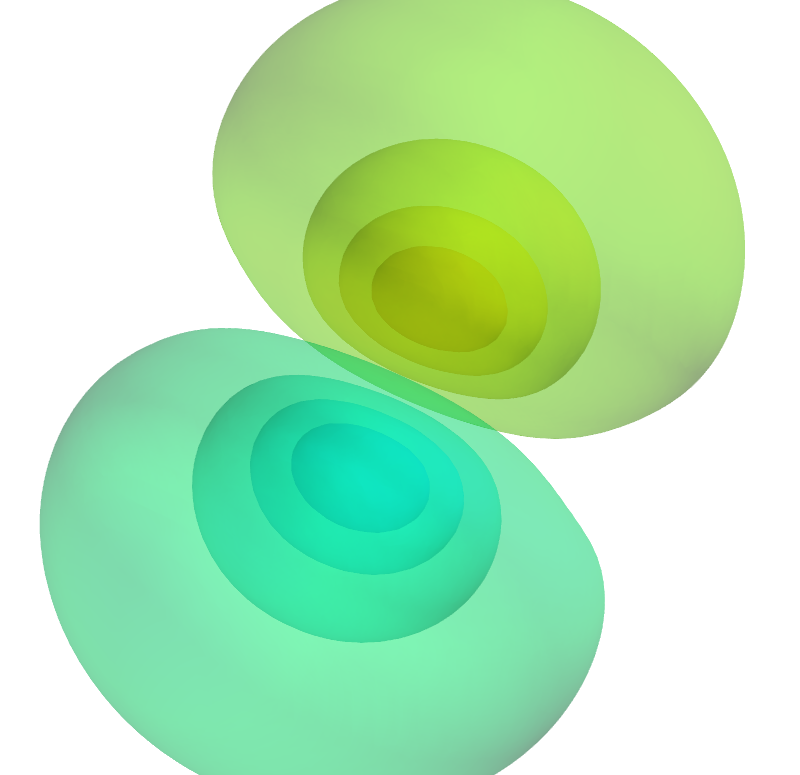}
    &
    \includegraphics[width=0.18\linewidth]{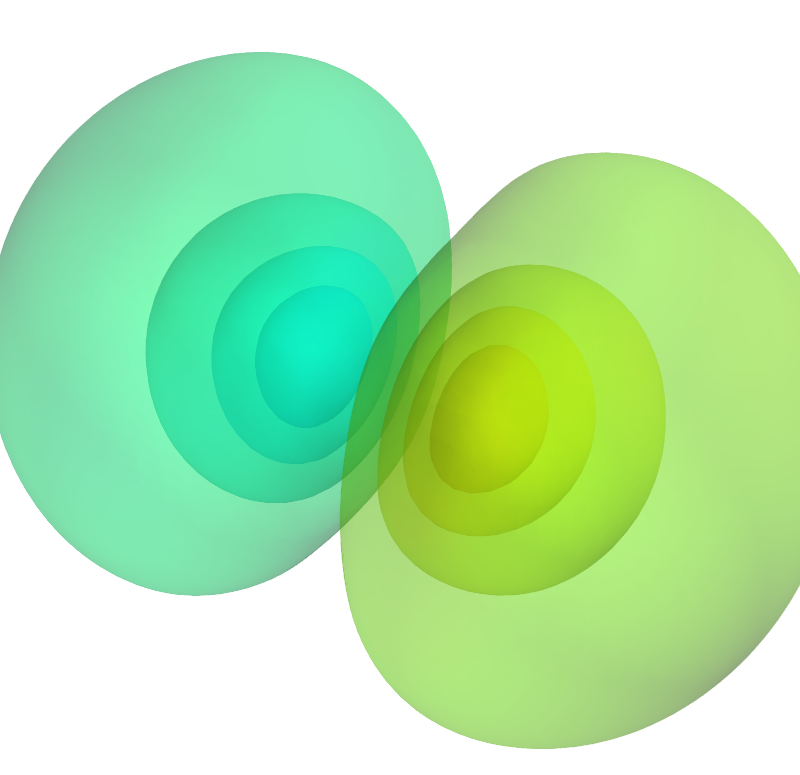}
    &
    \includegraphics[width=0.18\linewidth]{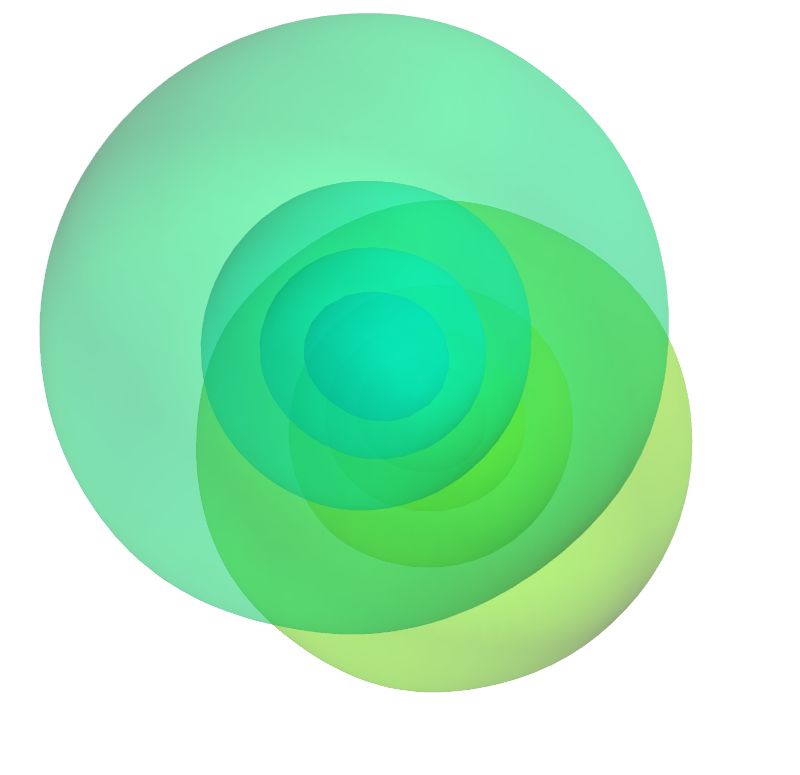}
    \\
    20 &
    \includegraphics[width=0.18\linewidth]{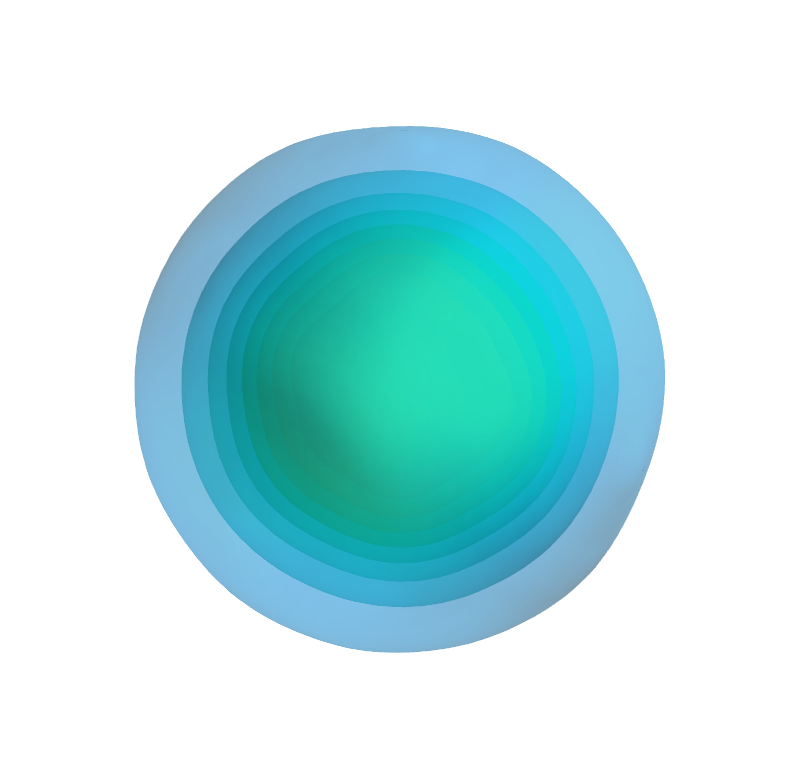}
    &
    \includegraphics[width=0.18\linewidth]{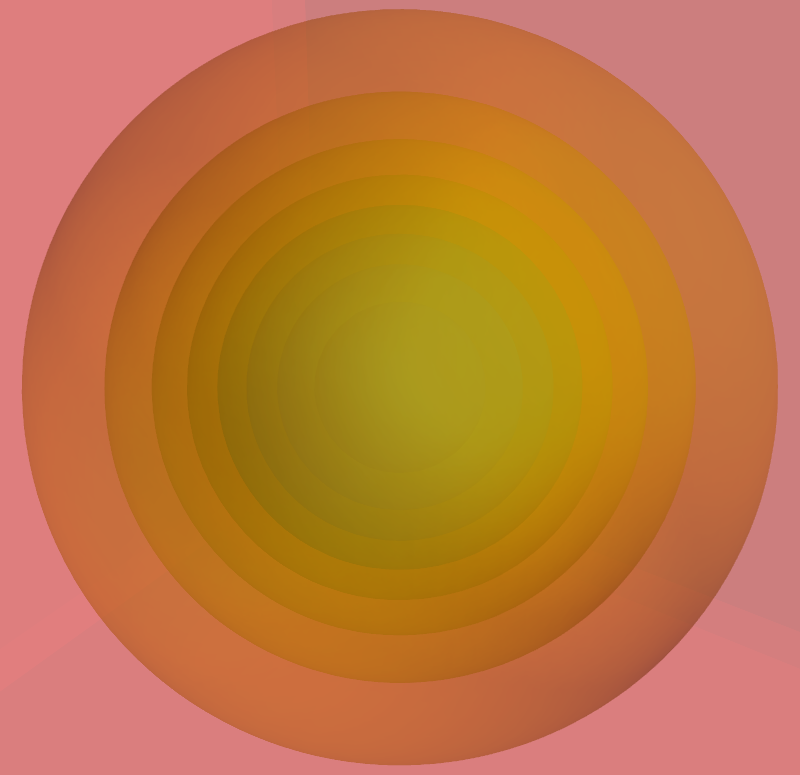}
    &
    \includegraphics[width=0.18\linewidth]{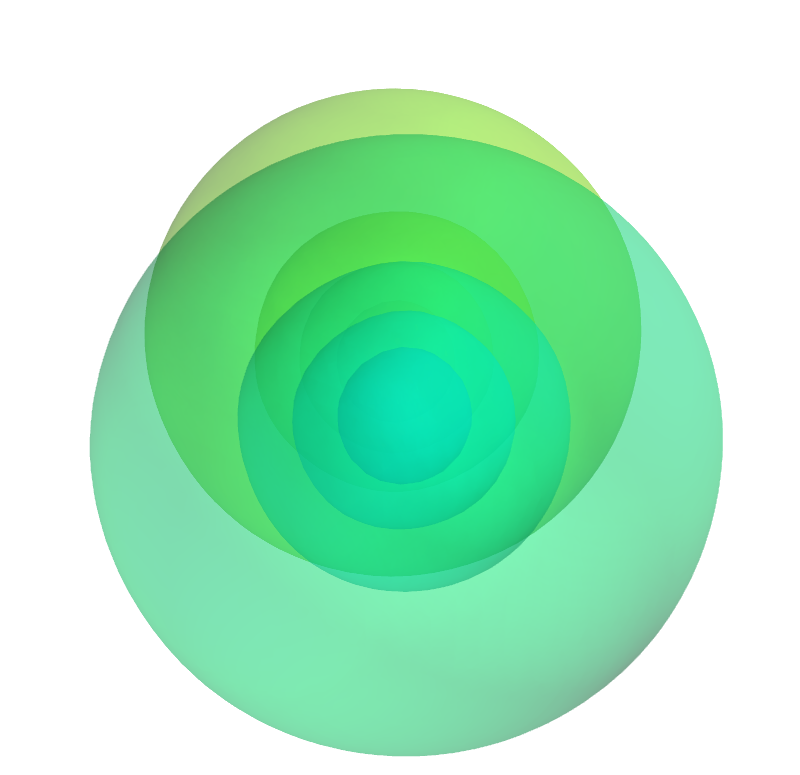}
    &
    \includegraphics[width=0.18\linewidth]{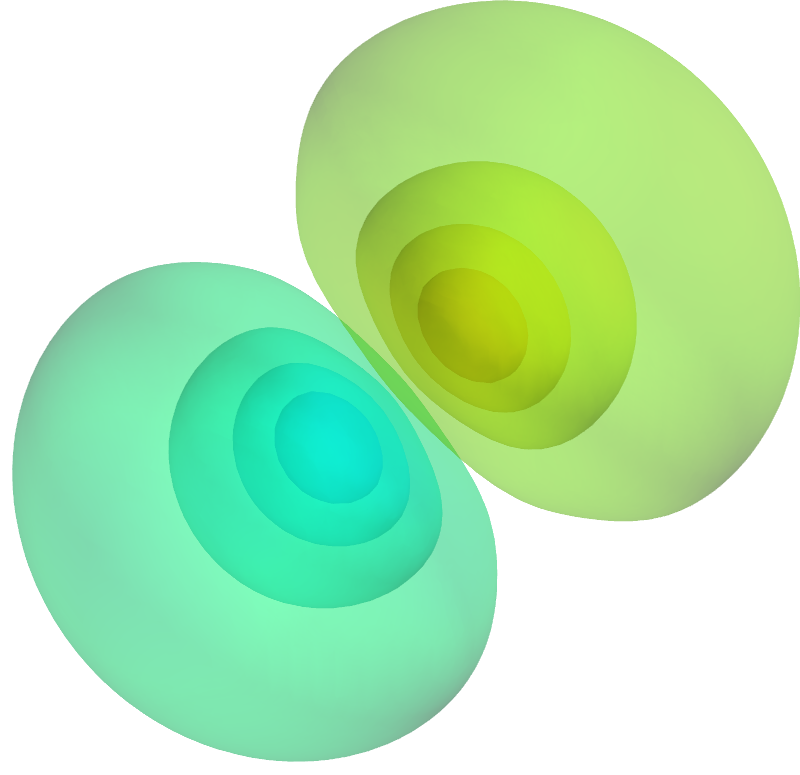}
    &
    \includegraphics[width=0.18\linewidth]{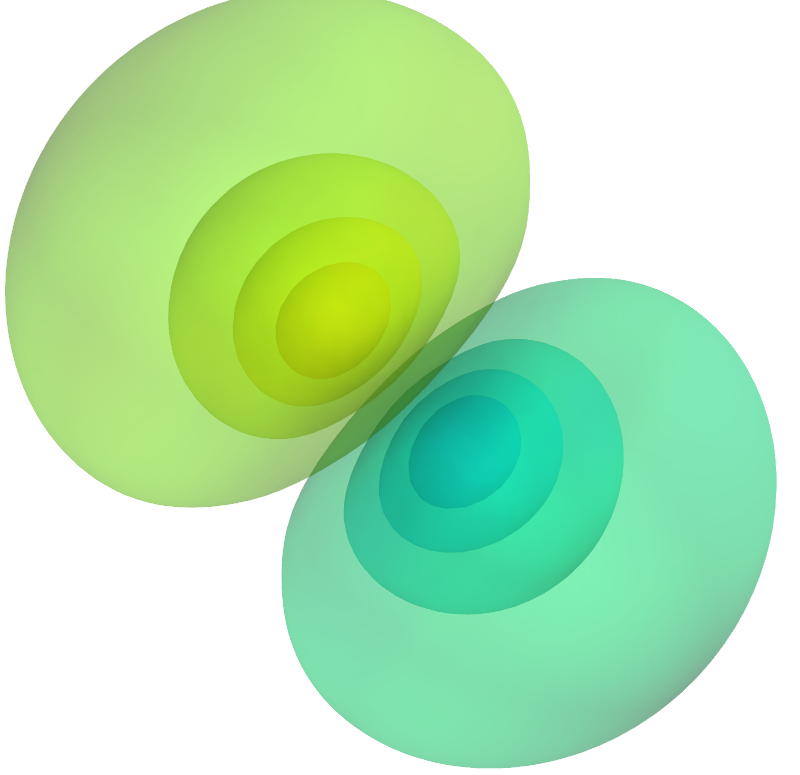}
    \\
    24 &
    \includegraphics[width=0.18\linewidth]{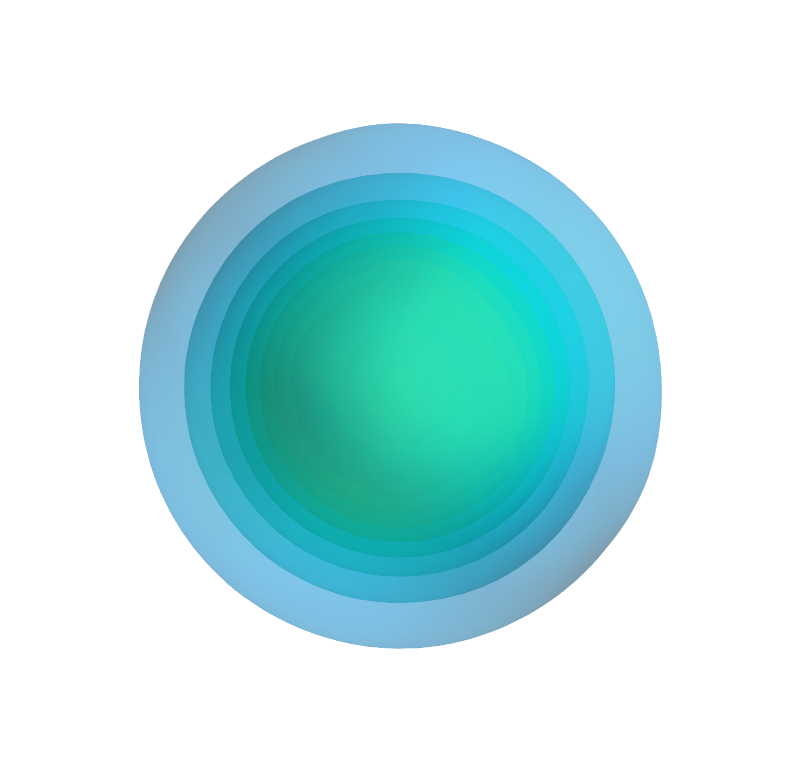}
    &
    \includegraphics[width=0.18\linewidth]{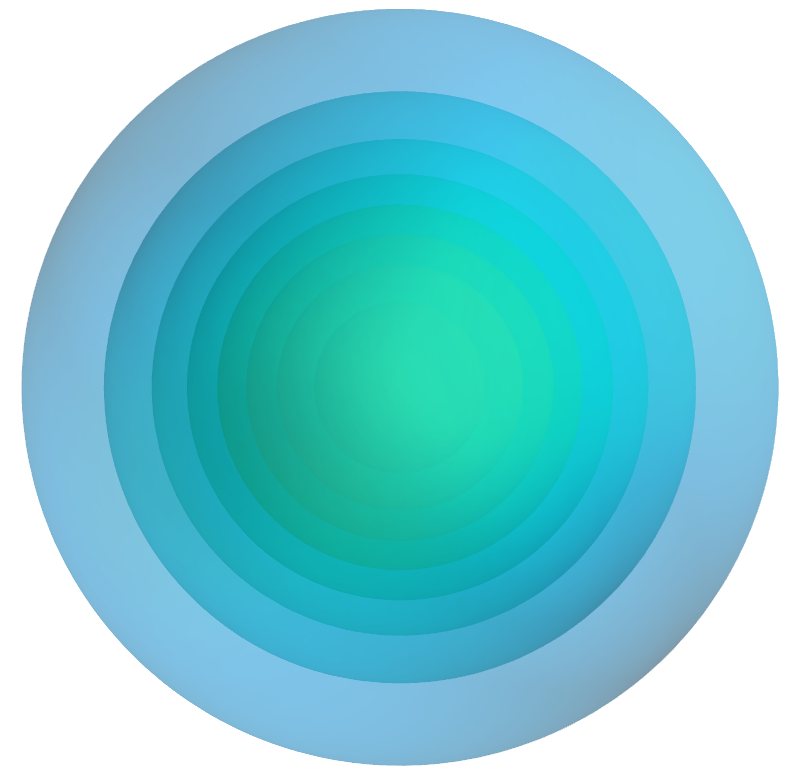}
    &
    \includegraphics[width=0.18\linewidth]{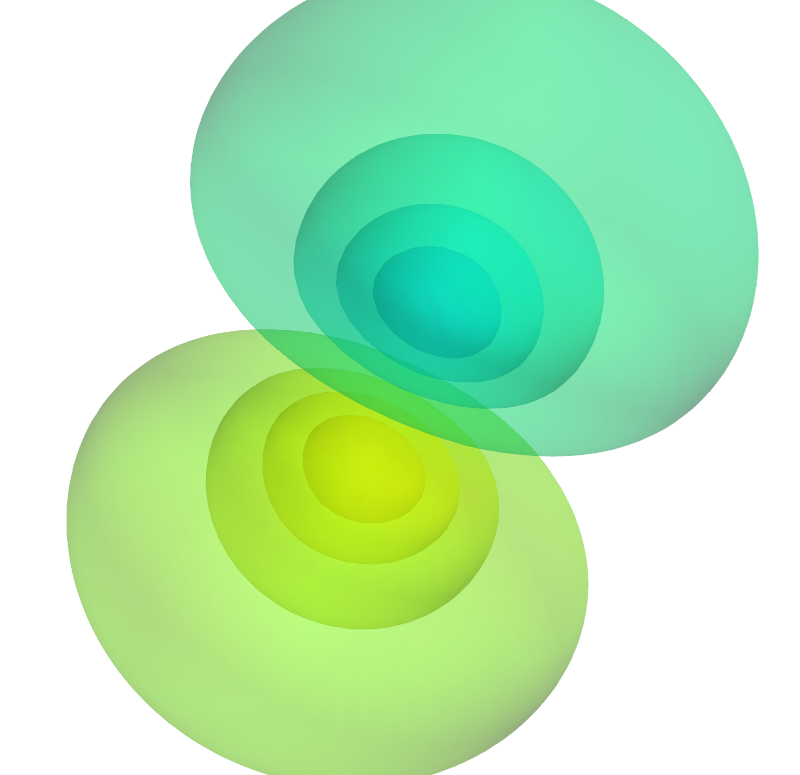}
    &
    \includegraphics[width=0.18\linewidth]{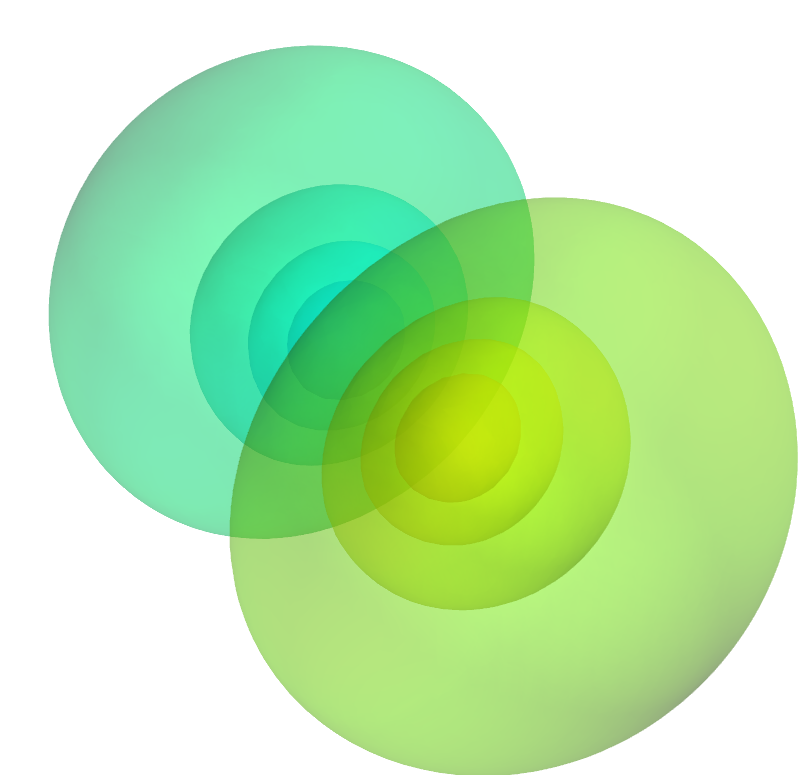}
    &
    \includegraphics[width=0.18\linewidth]{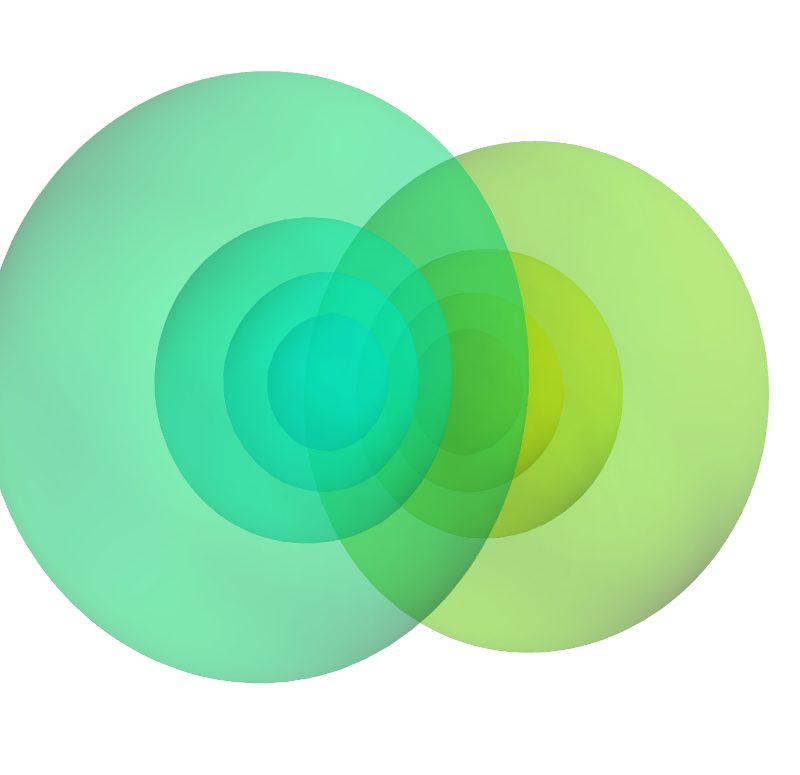}
    \\
    28 &
    \includegraphics[width=0.18\linewidth]{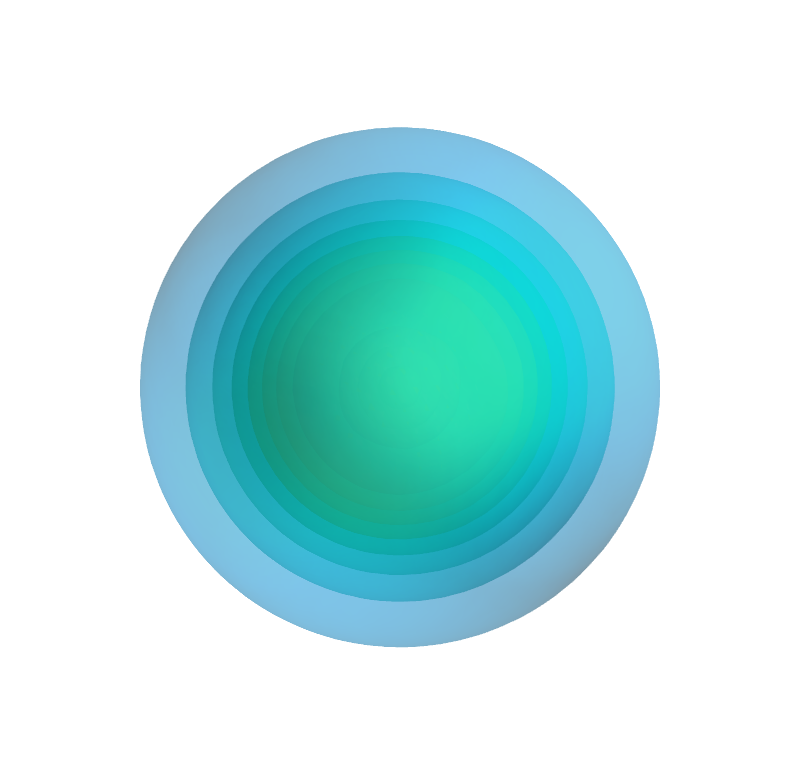}
    &
    \includegraphics[width=0.18\linewidth]{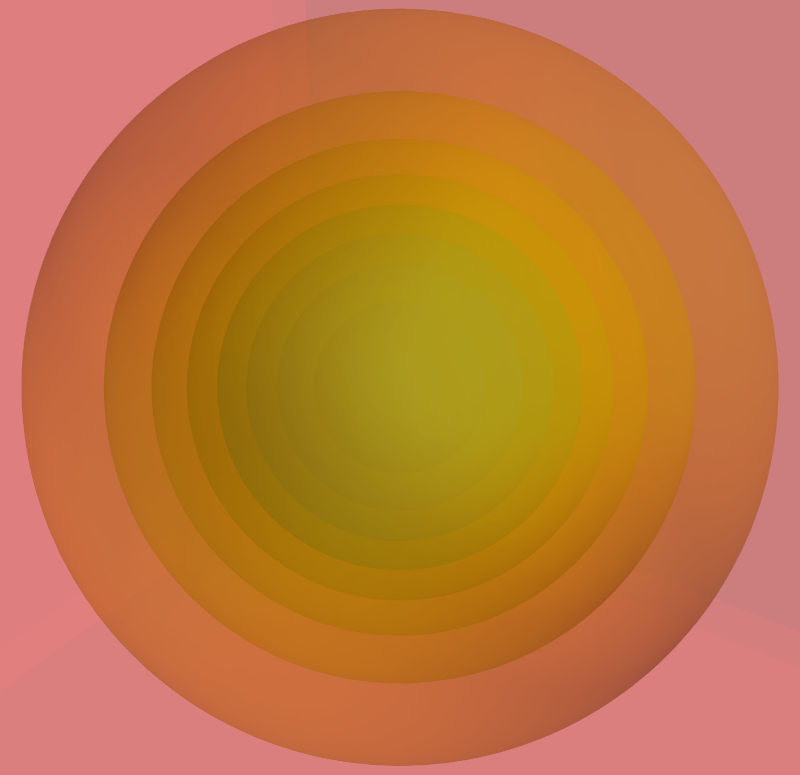}
    &
    \includegraphics[width=0.18\linewidth]{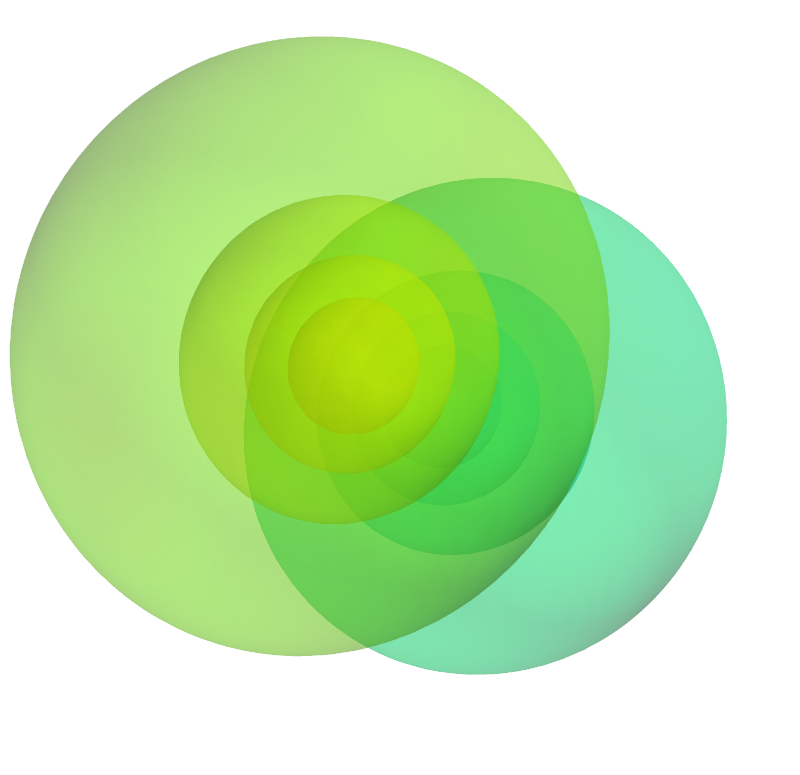}
    &
    \includegraphics[width=0.18\linewidth]{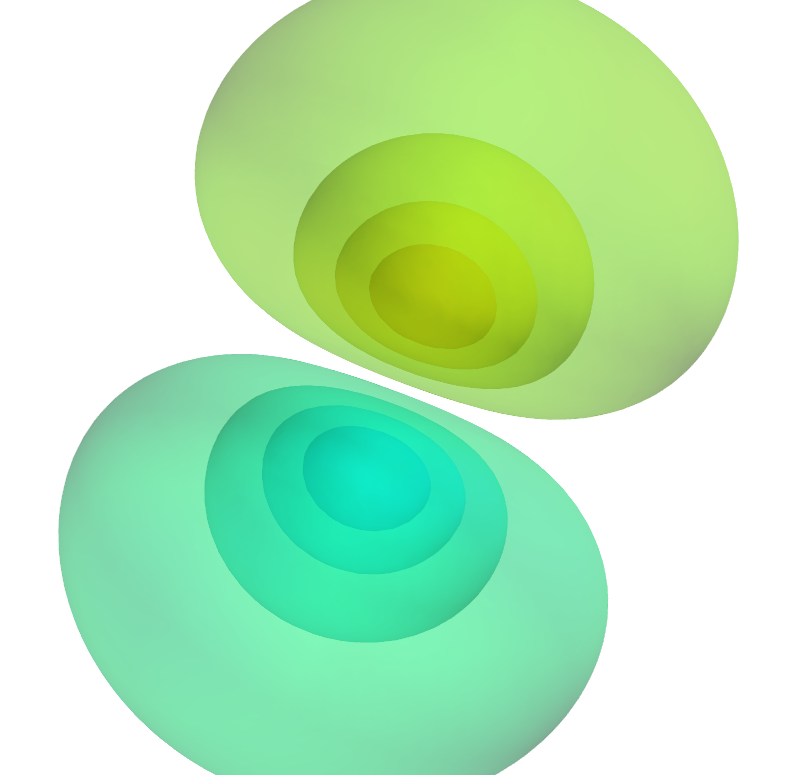}
    &
    \includegraphics[width=0.18\linewidth]{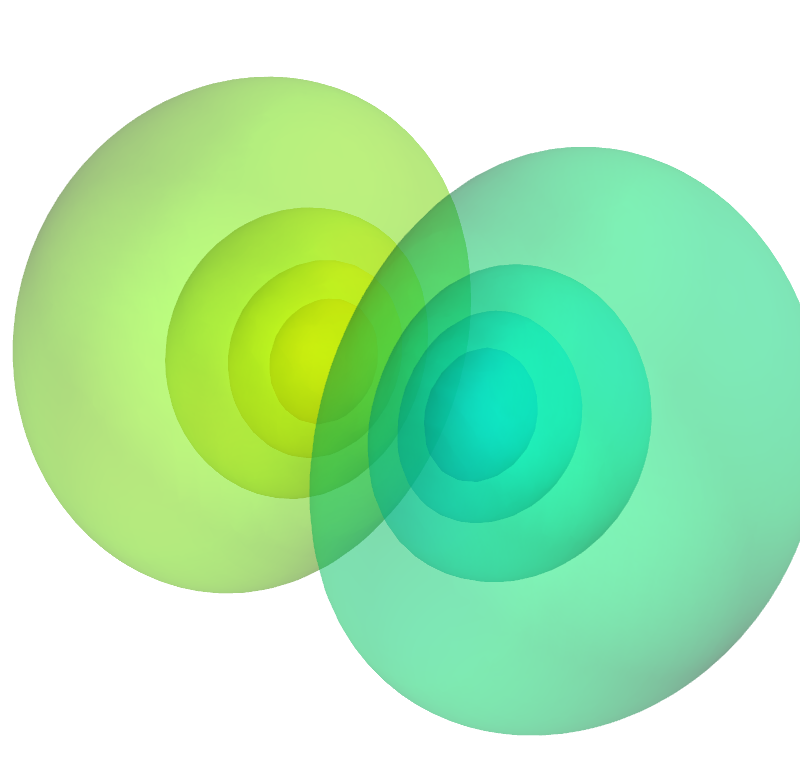}
    \\
    & $\rho$ & $\psi_1$ & $\psi_2$ & $\psi_3$ & $\psi_4$
  \end{tabular}

  \caption{Nitrogen atom: iso-surfaces of charge density $\rho$ and orbitals
    $\psi_i$ for several IGA grid sizes. The number of DOFs per axis is shown
    in the left-most column. The sizes of the individual images are in
    proportion. The color range shown in the top row is used in the
    corresponding column, except the $\psi_1$ plot for the 24 DOFs/axis grid
    --- there $\psi_1$ has the opposite sign then in the other rows. This is a
    side effect of the eigenvalue solver implementation and is physically
    insignificant.}
  \label{fig:atom}
\end{figure}

In physical simulations we are interested in other quantities, besides the
eigenvalues $\varepsilon_i$. As an example we show the charge density and the
orbitals $\psi_i$ of the nitrogen atom in Fig.~\ref{fig:atom} for some of the
grids used in the convergence study above. It shows that the electron states
form into shapes of spherical harmonics even if no preliminary
shape-anticipating assumption is done. Note that depending on the grid
resolution, the orientation of orbitals without spherical symmetry ($\psi_2$,
$\psi_3$, and $\psi_4$) changes.

\begin{figure}[htp!]
  \centering
  \includegraphics[width=0.48\linewidth]{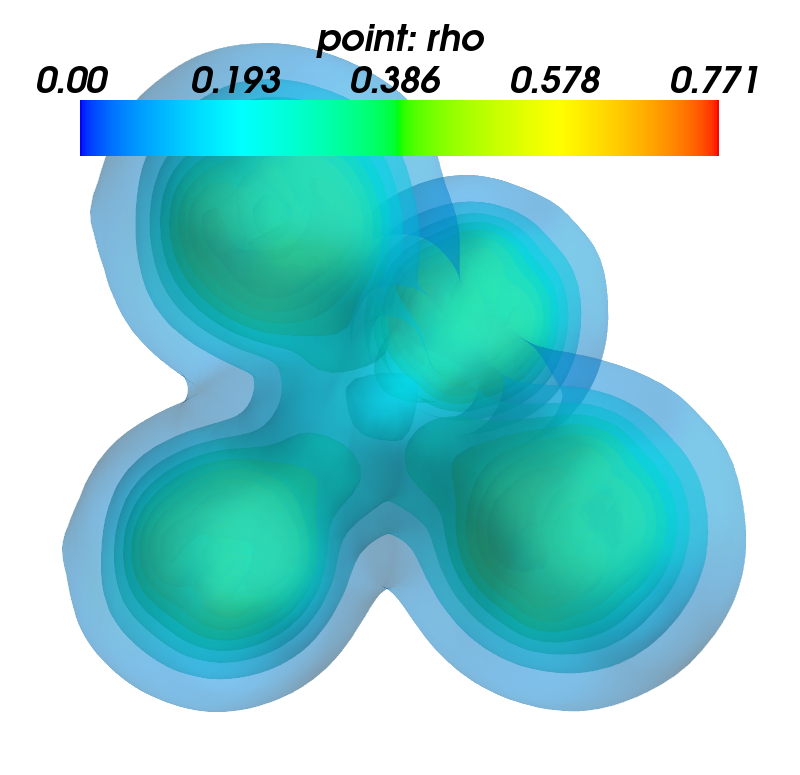}

  \caption{Tetrafluormethane molecule: iso-surfaces of charge density $\rho$.}
  \label{fig:cf4}
\end{figure}

For illustration of a more complex structure, we also present the distribution
of the charge density $\rho$ of the tetrafluormethane molecule (CF$_4$) in
Fig.~\ref{fig:cf4}. Even though the parametric grid used in this computation
was only $20 \times 20 \times 20$, with $C^2$ continuous B-spline basis, good
results were obtained: our code correctly reproduced both the angles of C-F
branches and the inter-atomic distances, by minimizing the total energy.

\section{Conclusion}

We introduced our approach to electronic structure calculations, based on the
density functional theory, the environment-reflecting pseudopotentials and a
weak solution of the Kohn-Sham equations. Our computer implementation built
upon the open source package SfePy supports computations both with the finite
element basis and the NURBS or B-splines basis of isogeometric analysis. The
latter allows a globally $C^2$ continuous approximation of unknown fields,
which is crucial for computing of derivatives of the total energy w.r.t. atomic
positions etc., as given by the Hellmann-Feynman theorem.

Numerical results comparing the FEM and IGA calculations were presented on the
benchmark problem of nitrogen atom. These results suggest significantly better
convergence properties of IGA over FEM for our application, due to the higher
smoothness of the approximation. This will be further studied on more complex
substances, together with the implementation of the calculation of total energy
derivatives. Finally, other quantities (charge density and related orbitals)
that can be computed were illustrated using figures.

To alleviate the numerical quadrature cost, reduced quadrature rules has been
proposed for the context of the Bézier extraction \cite{IGA-Schillinger-2},
which we plan to assess in future.

\noindent
\textbf{Acknowledgments:} The work was supported by the Grant Agency of the
Czech Republic, project P108/11/0853. R. Kolman's work was supported by the
grant project of the Czech Science Foundation (GACR), No. GAP 101/12/2315,
within the institutional support RVO:61388998.

\section*{References}

\bibliography{modelling2014-iga-dft.bib}

\begin{thebibliography}{10}
\expandafter\ifx\csname url\endcsname\relax
  \def\url#1{\texttt{#1}}\fi
\expandafter\ifx\csname urlprefix\endcsname\relax\def\urlprefix{URL }\fi

\bibitem{IGA-6}
Y.~Bazilevs, V.~M. Calo, J.~A. Cottrell, J.~A. Evans, T.~J.~R. Hughes,
  S.~Lipton, M.~A. Scott, T.~W. Sederberg, Isogeometric analysis using
  {T}-splines, Comput. Methods Appl. Mech. Engrg. 199 (2010) 229--263.

\bibitem{IGA-2}
M.~J. Borden, M.~A. Scott, J.~A. Evans, T.~J.~R. Hughes, Isogeometric finite
  element data structures based on {B}ezier extraction of {NURBS}, Int. J.
  Numer. Meth. Engng. 87 (2011) 15--47.

\bibitem{SfePy-2}
R.~Cimrman, Enhancing {SfePy} with isogeometric analysis, in: P.~de~Buyl,
  N.~Varoquaux (eds.), Proceedings of the 7th European Conference on Python in
  Science (EuroSciPy 2014), 2014, pp. 65--72.
\newline\urlprefix\url{http://arxiv.org/abs/1412.6407}

\bibitem{SfePy-1}
R.~Cimrman, {SfePy} - write your own {FE} application, in: P.~de~Buyl,
  N.~Varoquaux (eds.), Proceedings of the 6th European Conference on Python in
  Science (EuroSciPy 2013), 2014, pp. 65--70.
\newline\urlprefix\url{http://arxiv.org/abs/1404.6391}

\bibitem{IGA-Collier}
N.~Collier, D.~Pardo, L.~Dalcin, M.~Paszynski, V.~M. Calo, The cost of
  continuity: A study of the performance of isogeometric finite elements using
  direct solvers, Computer Methods in Applied Mechanics and Engineering
  213–216 (2012) 353 -- 361.

\bibitem{IGA-1}
J.~A. Cottrell, T.~J.~R. Hughes, Y.~Bazilevs, Isogeometric Analysis: Toward
  Integration of CAD and FEA, John Wiley \& Sons, Chichester, West Sussex,
  U.K., 2009.

\bibitem{IGA-3}
J.~A. Cottrell, T.~J.~R. Hughes, A.~Reali, Studies of refinement and continuity
  in isogeometric structural analysis, Comput. Methods Appl. Mech. Engrg. 196
  (2007) 4160--4183.

\bibitem{DFT-FEM-Davydov}
D.~Davydov, T.~D. Young, P.~Steinmann, On the adaptive finite element analysis
  of the {K}ohn-{S}ham equations: Methods, algorithms, and implementation,
  International Journal for Numerical Methods in Engineering (2015) n/a--n/a.

\bibitem{Alessandra}
A.~Di~Pomponio, A.~Continenza, R.~Podloucky, J.~Vackář, Symmetrization of
  atomic forces within the full-potential linearized augmented-plane-wave
  method, Phys. Rev. B 53 (1996) 9505--9508.

\bibitem{DFT-1}
R.~M. Dreizler, E.~K.~U. Gross, Density Functional Theory, Springer-Verlag,
  Berlin, 1990.

\bibitem{FEM-1}
T.~J.~R. Hughes, The Finite Element Method: Linear Static and Dynamic Finite
  Element Analysis, Dover Publications, Mineola, New York, USA, 2000.

\bibitem{IGA-5}
T.~J.~R. Hughes, J.~A. Evans, A.~Reali, Finite element and {NURBS}
  approximations of eigenvalue, boundary-value, and initial-value problems,
  Comput. Methods Appl. Mech. Engrg. 272 (2014) 290--320.

\bibitem{IGA-Waves-Hughes}
T.~J.~R. Hughes, A.~Reali, G.~Sangalli, Duality and unified analysis of
  discrete approximations in structural dynamics and wave propagation:
  Comparison of p-method finite elements with k-method {NURBS}, Computer
  Methods in Applied Mechanics and Engineering 197~(49–50) (2008) 4104--4124.

\bibitem{IhmZungerCohen}
J.~Ihm, A.~Zunger, C.~M. L., Momentum-space formalism for the total energy of
  solids, J. Phys. C: Solid State Phys. 12 (1979) 4409--4422.

\bibitem{Kohn-Sham}
W.~Kohn, L.~J. Sham, Self-consistent equations including exchange and
  correlation effects, Phys. Rev. 140~(4A) (1965) A1133–A1138.

\bibitem{IGA-4}
R.~Kolman, J.~Ple\v{s}ek, M.~Okrouhl\'\i{}k, Complex wavenumber {F}ourier
  analysis of the {B}-spline based finite element method, Wave Motion 51 (2013)
  348--359.

\bibitem{IGA-8}
R.~Kolman, S.~V. Sorokin, B.~Bastl, J.~Kopa\v{c}ka, J.~Ple\v{s}ek, Isogeometric
  analysis of free vibration of simple shaped elastic samples, Journal of the
  Acoustical Society of America 137~(4) (2015) 2089--2100.

\bibitem{DFT-3}
R.~M. Martin, Electronic Structure: Basic Theory and Practical Methods,
  Cambridge University Press, Cambridge, New York, USA, 2005.

\bibitem{DFT-IGA-Masud}
A.~Masud, R.~Kannan, B-splines and \{NURBS\} based finite element methods for
  {K}ohn–{S}ham equations, Computer Methods in Applied Mechanics and
  Engineering 241–244 (2012) 112--127.

\bibitem{DFT-FEM-Motamarri-1}
P.~Motamarri, V.~Gavini, Subquadratic-scaling subspace projection method for
  large-scale {K}ohn-{S}ham density functional theory calculations using
  spectral finite-element discretization, Phys. Rev. B 90 (2014) 115127.

\bibitem{DFT-FEM-Motamarri-2}
P.~Motamarri, M.~R. Nowak, K.~Leiter, J.~Knap, V.~Gavini, Higher-order adaptive
  finite-element methods for {K}ohn–{S}ham density functional theory, Journal
  of Computational Physics 253 (2013) 308--343.

\bibitem{DFT-2}
R.~G. Parr, Y.~Weitao, Density-Functional Theory of Atoms and Molecules, Oxford
  University Press, Oxford, USA, 1994.

\bibitem{DFT-4}
W.~E. Pickett, Pseudopotential methods in condensed matter applications, Comp.
  Phys. Reports 9 (1989) 115--198.

\bibitem{NURBS}
L.~Piegl, W.~Tiller, {T}he {NURBS} {B}ook, (2nd ed.) ed., Springer-Verlag, New
  York, New York, USA, 1995–1997.

\bibitem{IGA-Schillinger-1}
D.~Schillinger, J.~A. Evans, A.~Reali, M.~A. Scott, T.~J.~R. Hughes,
  Isogeometric collocation: Cost comparison with {G}alerkin methods and
  extension to adaptive hierarchical \{NURBS\} discretizations, Computer
  Methods in Applied Mechanics and Engineering 267 (2013) 170 -- 232.

\bibitem{IGA-Schillinger-2}
D.~Schillinger, S.~J. Hossain, T.~J.~R. Hughes, Reduced {B}ézier element
  quadrature rules for quadratic and cubic splines in isogeometric analysis,
  Computer Methods in Applied Mechanics and Engineering 277 (2014) 1--45.

\bibitem{FEM-2}
G.~Strang, G.~Fix, An Analysis of the Finite Element Method,
  Wellesley-Cambridge Press, Wellesley, 2008, pp. 414.

\bibitem{vackar}
J.~Vack\'{a}\v{r}, A.~\v{S}im\accent23 unek, Adaptability and accuracy of
  all-electron pseudopotentials, Phys. Rev. B 67 (2003) 125113.

\bibitem{ptcp}
J.~Vackář, O.~Čertík, R.~Cimrman, M.~Novák, O.~Šipr, J.~Plešek, Advances
  in the Theory of Quantum Systems in Chemistry and Physics, vol.~22 of Prog.
  Theoretical Chem. and Phys., chap. Finite Element Method in Density
  Functional Theory Electronic Structure Calculations, Springer Netherlands,
  Netherlands, 2011, pp. 199--217.

\bibitem{Weinert}
M.~Weinert, J.~W. Davenport, Fractional occupations and density-functional
  energies and forces, Phys. Rev. B 45 (1992) 13709--13712.

\bibitem{Krakauer}
R.~Yu, D.~Singh, H.~Krakauer, All-electron and pseudopotential force
  calculations using the linearized-augmented-plane-wave method, Phys. Rev. B
  43 (1991) 6411--6422.

\end{thebibliography}

\end{document}